\documentclass[5p]{elsarticle}
\makeatletter
\def\ps@pprintTitle{%
 \let\@oddhead\@empty
 \let\@evenhead\@empty
 \def\@oddfoot{}%
 \let\@evenfoot\@oddfoot}
\makeatother

\usepackage{csquotes}
\usepackage[dvipsnames]{xcolor}
\usepackage{graphicx,color}
\usepackage{float}
\usepackage[caption=false]{subfig}
\usepackage{amsmath}
\usepackage{amssymb}
\usepackage{multirow}
\usepackage{dsfont}
\usepackage{adjustbox}
\usepackage{pdflscape}
\usepackage{wasysym}
\usepackage{textcomp}
\usepackage{enumitem}
\usepackage{stackengine}
\usepackage{soul}
\usepackage[percent]{overpic}
\def\mathcolor#1#{\@mathcolor{#1}}
\def\@mathcolor#1#2#3{%
  \protect\leavevmode
  \begingroup
    \color#1{#2}#3%
  \endgroup
}
\usepackage{tikz}
\usetikzlibrary{shapes.geometric, arrows}
\tikzstyle{startstop} = [rectangle, rounded corners, minimum width=2cm, minimum height=1cm,text centered, draw=black, fill=gray!30]
\tikzstyle{io} = [trapezium, trapezium left angle=70, trapezium right angle=110, minimum width=3cm, minimum height=1cm, text centered, draw=black, fill=blue!30]
\tikzstyle{process} = [rectangle, minimum width=4cm, minimum height=1cm, text centered, text width=4cm, draw=black, fill=gray!30]
\tikzstyle{process_mini} = [rectangle, minimum width=1.5cm, minimum height=1cm, text centered, text width=1.5cm, draw=black, fill=gray!30]
\tikzstyle{recursive} = [rectangle, minimum width=4cm, minimum height=1cm, text centered, text width=4cm, draw=black, fill=yellow!60]
\tikzstyle{decision} = [diamond, minimum width=2cm, minimum height=0.5cm, text centered, draw=black, fill=gray!30]
\tikzstyle{arrow} = [thick,->,>=stealth]
\usetikzlibrary{automata, positioning}
\tikzstyle{arrowc} = [thick,-->,>=stealth]

\begin{document}
\title{The MEMENTO code for modelling of macroscopic melt motion in fusion devices}
\author{K. Paschalidis$^{a}$, F. Lucco Castello$^{a}$, S. Ratynskaia$^{a}$, P. Tolias$^{a}$, and L. Brandt$^{b,c}$}
\address{
$^a$ Space and Plasma Physics, KTH Royal Institute of Technology, Stockholm SE-100 44, Sweden \\
$^b$ Department of Energy and Process Engineering, Norwegian Universtity of Science and Technology (NTNU), Trondheim, Norway \\
$^c$ FLOW, Department of Engineering Mechanics, KTH Royal Institute of Technology, Stockholm SE-100 44, Sweden
}
% \ead{\mailto{srat@kth.se}}
\begin{abstract}
The {\bf MEMENTO} ({\bf ME}tallic {\bf M}elt {\bf E}volution in {\bf N}ext-step {\bf TO}kamaks) code is a new numerical implementation of the physics model originally developed for the MEMOS-U code with the objective to self-consistently describe the generation of melt and its subsequent large scale dynamics in fusion devices and to assess the damage of metallic reactor armor under powerful normal and off-normal plasma events. The model has been validated in multiple dedicated EUROfusion experiments. MEMENTO solves the heat and phase transfer problem coupled with the incompressible Navier-Stokes equations in the shallow water approximation for the thin liquid film over the solid metal and with the current propagation equations on a domain that features a time-evolving deforming metal-plasma interface. The code utilizes non-uniform and adaptive meshing along with sub-cycling in time facilitated by the AMReX open-source framework as well as AMReX's built-in parallelization capabilities.
\end{abstract}
\maketitle

\section{Introduction}

Power handling and plasma-facing component (PFC) longevity constitute major technological obstacles on the way to efficient operation of fusion reactors~\cite{Pitts2017}. The melting of metallic armor under high energy-density transient events is the main threat to PFC integrity, since liquid metal displaced by plasma-induced or external forces can cause large-scale surface deformations and even modify the castellated PFC structure. Since contemporary fusion devices cannot achieve the plasma parameters of future machines, such as ITER~\cite{Pitts2017} and DEMO~\cite{Maviglia2022}, predictive modelling of thermal response and melt motion is crucial in guiding reactor designs and choices of safe operational space.

The MEMOS-U physics model~\cite{Ratynskaia2020,Ratynskaia2021,Thoren2021,Thoren2018}, addressing the formation and dynamics of metallic melt pools in fusion devices, has been developed over several years in a coordinated theoretical, computational and experimental effort within the EUROfusion framework~\cite{Corre2023,Corre2021,Krieger2017,Krieger2018,Coenen2015,Jepu2019,Ratynskaia2022_1,Ratynskaia2023}. The relevant scenarios concern multi-phase (liquid metal, metal vapor, plasma) magneto-hydrodynamic flows with highly deformed free-surfaces and dynamic bathymetry due to the propagation of the melting and/or solidification fronts. Given the vast scale separation between the pool extent and depth, the shallow water approximation is applicable~\cite{Acheson1990, Vreugdenhil1994}. This allows the introduction of plasma-induced effects through boundary conditions on the free-surface~\cite{Ratynskaia2021,Thoren2021,Thoren2018}, some of which are based on semi-empirical scalings devised from dedicated simulations of magnetized multi-emissive plasma sheaths~\cite{Komm2017,Komm2017_1,Komm2020,Tolias2020,Tolias2023}.

The physics model and treatment of plasma-related effects have been guided by empirical evidence from a series of dedicated tokamak experiments~\cite{Corre2021,Krieger2017,Krieger2018,Coenen2015,Jepu2019,Ratynskaia2022_1}. The numerical implementation of the  model in the eponymous code~\cite{Ratynskaia2021,Thoren2021,Thoren2018} has been correspondingly carried out in an additive fashion. Moreover, as validation against experiments proved successful, the use of the tool for predictive modelling for fusion reactors became essential. Such applications pose additional challenges due to the necessity to simulate intricate PFC layouts as well as complex plasma-wetted geometries. Furthermore, simulations of at least an entire plasma-facing unit (monoblock, castellation) are necessary, where it is essential to resolve the thin localized melt pools on timescales that include long power loading combined with short quasi-periodic bursts due to edge localized modes (ELMs). The lack of a priori knowledge of the PFC geometries and wetting details as well as the spatiotemporal scales to be tackled inhibited the development of efficient numerical solutions. The original implementation of the MEMOS-U physics model was computationally inefficient and lacked flexibility in simulations of complex domains and complicated scenarios. This served as the primary incentive behind the development of a new implementation of the physics model in a modern computational tool - MEMENTO ({\bf ME}tallic {\bf M}elt {\bf E}volution in {\bf N}ext-step {\bf TO}kamaks).

In this work, an in-depth overview of MEMENTO is presented, including a detailed description of the adopted numerical schemes. The MEMENTO code relies on AMReX, a software framework that provides functionality for massively parallel codes and adaptive meshing techniques~\cite{Amrex, Zhang2019_amrex}. Numerical benchmarks of the solvers of MEMENTO have been presented in a previous publication~\cite{Paschalidis2023}. Here, \textcolor{red}the code structure is discussed and a simulation flowchart is provided together with distinctive examples of the MEMENTO capabilities concerning the modelling of fusion-relevant melting events.

\section{Model equations}

\begin{figure}
    \includegraphics[width=\columnwidth]{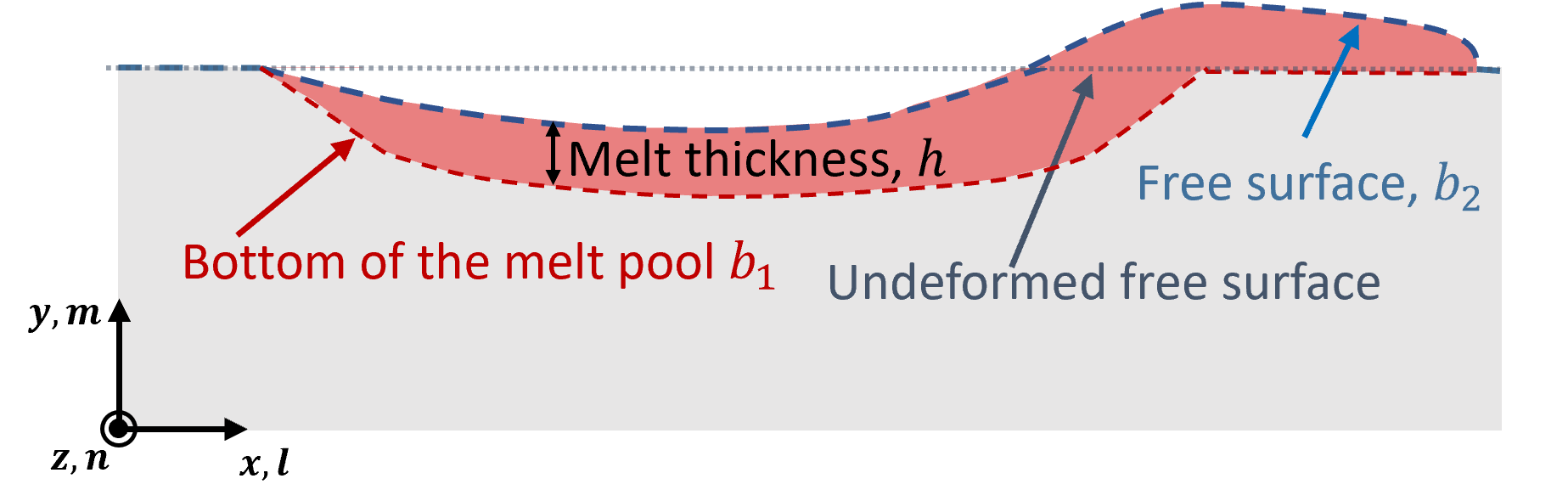}
    \caption{ Sketch of a typical MEMENTO simulation scenario. The cartesian coordinate system $(x,y,z)$ has been assigned with the indices $(l,m,n)$ that are employed in the discretized form of equations. The melt pool bottom, $b_1$, the initial and deformed free surface $b_2$, and the melt thickness, $h$, are also illustrated.}
    \label{fig:MEMENTO_coordinate_sketch}
\end{figure}

A typical scenario of concern to fusion-relevant melting events involves an initially solid metal component subject to a type of plasma heat load on its surface for a given duration. Once a melt pool is formed, it is subject to plasma-induced forces which accelerate the liquid, displacing it from its original position. Since the plasma heat loads can exhibit strong spatiotemporal variations, the pools can be transient and feature a deforming free-surface as well as an evolving bathymetry due to the propagation of the melting or resolification fronts. A sketch of a typical domain on which the model's equations are solved is illustrated in Fig.\ref{fig:MEMENTO_coordinate_sketch}. The MEMENTO coordinate system is introduced together with the indices utilized for the spatial directions when discussing the discretized form of the equations. The index $i$ is reserved for the temporal variations. This convention is respected throughout the text.

The starting point for the modelling of fusion melting events concerns the thermo-electric magnetohydrodynamic equations (TEMHD) coupled with the heat conduction equation, as formulated by Sherclif~\cite{Shercliff1979}.  The implemented physics model~\cite{Ratynskaia2020, Ratynskaia2021,Thoren2021,Thoren2018_1} is based on the TEMHD equations with the electromagnetic field equations formulated within the magnetostatic limit for a uniform material composition and the Navier-Stokes equations formulated within the shallow water (SW) approximation.

In what follows, for the sake of completeness, the set of equations is recapped, while all the physical quantities and thermophysical properties involved are introduced:
\begin{align}
    \label{eqn:column_height}
    &\frac{\partial h}{\partial{t}}+\nabla_{\mathrm{t}}\cdot(h\boldsymbol{U})=-\frac{\partial{b}_{1}}{\partial{t}}-\dot{x}_{\mathrm{vap}},
    \\
    \label{eqn:momentum}
    &\rho_{\mathrm{m}}\left[\frac{\partial \boldsymbol{U}}{\partial{t}}+\left(\boldsymbol{U}\cdot\nabla_{\mathrm{t}}\right)\boldsymbol{U}\right] = \langle\boldsymbol{J}\rangle\times\boldsymbol{B}-\frac{3\mu}{h^{2}}\boldsymbol{U}+\mu\nabla_{\mathrm{t}}^{2}\boldsymbol{U}+\nonumber\\
    & \hspace{0.5cm}\rho_{\mathrm{m}}\boldsymbol{g}+\frac{3}{2h}\left(\frac{\partial\sigma}{\partial{T}}\nabla_{\mathrm{t}}T_{\mathrm{s}}+\boldsymbol{f}_{\mathrm{d}}\right)-\nabla_{\mathrm{t}}P + \sigma\nabla_{\mathrm{t}}\nabla_{\mathrm{t}}^2b_2,
    \\
    \label{eqn:energy}
    &c_{\mathrm{p}}\rho_{\mathrm{m}}\left[\frac{\partial{T}}{\partial{t}}+\boldsymbol{U}\cdot\nabla_{\mathrm{t}}T\right]=\nabla\cdot(k\nabla{T})+V,
    \\
    \label{eqn:potential}
    &\nabla\cdot\left[\frac{1}{\rho_{\mathrm{e}}}\nabla\psi\right]=\nabla \cdot \left[ \frac{1}{\rho_{\mathrm{e}}} {\boldsymbol{U}} \times \boldsymbol{B} \right],
    \\
    \label{eqn:current}
    &\boldsymbol{J}=-\frac{1}{\rho_{\mathrm{e}}}\left(\nabla\psi-\boldsymbol{U}\times\boldsymbol{B}\right),
\end{align}
with the corresponding boundary conditions (BCs)
\begin{align}
    \label{eqn:no_inflow}
    \tag{1a}
    &(h\boldsymbol{U})\cdot\widehat{\boldsymbol{n}}_{\mathrm{t}}=0\text{ if }\boldsymbol{U}\cdot\widehat{\boldsymbol{n}}_{\mathrm{t}}<0,
    \\
    \label{eqn:free_outflow}
    \tag{2a}
    &\left(\widehat{\boldsymbol{n}}_{\mathrm{t}}\cdot\nabla_{\mathrm{t}}\right)\boldsymbol{U}=0,
    \\
    \label{eqn:heat_BC}
    \tag{3a}
    % &-\left.k \nabla T\right|_{EB}=q_{\mathrm{inc}}-q_{\mathrm{cool}},\\
    &-k\frac{\partial{T}\left(\boldsymbol{r}\right)}{\partial{n}_{\mathrm{b}}}=q_{\mathrm{ext}}\left(\boldsymbol{r}\right),\forall\boldsymbol{r}\in{S}_{\mathrm{b}},
    \\
    \label{eqn:fs_charge_conservation}
    \tag{4a}
    &\frac{\partial\psi(\boldsymbol{r})}{\partial{n}_{\mathrm{b}_2}}=-\rho_{\mathrm{e}}J_{\mathrm{em}}(\boldsymbol{r}) + \left( {\boldsymbol{U}} \times \boldsymbol{B} \right) \cdot {\boldsymbol{\hat{n}}}_{\mathrm{b}_2},\forall\boldsymbol{r}\in{{b}_2},
    \\
    \label{eqn:insulation}
    \tag{4b}
    &\frac{\partial\psi(\boldsymbol{r})}{\partial{n}_{\mathrm{i}}} = \left( {\boldsymbol{U}} \times \boldsymbol{B} \right) \cdot {\boldsymbol{\hat{n}}}_{\mathrm{i}},\forall\boldsymbol{r}\in{S}_{\mathrm{i}},
    \\
    \label{eqn:ground}
    \tag{4c}
    &\psi(\boldsymbol{r})=0,\forall\boldsymbol{r}\in{S}_{\mathrm{g}}.
\end{align}

In the above, the liquid column height $h$ is defined as $h=b_2-b_1$, with $b_1$ and $b_2$ denoting the evolving surface of the liquid-solid interface and the evolving metal-plasma interface (free surface) respectively, as sketched in Fig.\ref{fig:MEMENTO_coordinate_sketch}. \emph{The SW equations} \eqref{eqn:column_height}-\eqref{eqn:momentum} and their BCs \eqref{eqn:no_inflow}, \eqref{eqn:free_outflow} describe the evolution of the film thickness $h$ and the depth-averaged velocity $\boldsymbol{U}$; $\dot{x}_{\mathrm{vap}}$ is the rate of change of $b_2$ due to vaporization, $\boldsymbol{\hat{n}}_t$ is a unit vector tangential to the free surface, $\boldsymbol{J}$ the bulk current density whose depth-averaged value is defined as
\begin{equation}
    \label{eqn:current_analytical_average}
    \langle \boldsymbol{J} \rangle = \frac{1}{h}\int_{b_1}^{b_2} \boldsymbol{J} dy,
\end{equation}
$\boldsymbol{B}$ the external magnetic field, $\rho_{\mathrm{m}}$ the mass density, $\sigma$ the surface tension, $\boldsymbol{f}_{\mathrm{d}}$ the drag force and $P$ the ambient pressure applied by the plasma and vapour at the surface. \emph{In the heat equation} \eqref{eqn:energy} and its BC \eqref{eqn:heat_BC}; $c_{\mathrm{p}}$ is the specific isobaric heat capacity, $k$ the thermal conductivity, $V$ the volumetric heat sources, $S_{\mathrm{b}}$ any bound of the computational domain for which Eq.(\ref{eqn:energy}) is solved, $\boldsymbol{\hat{n}}_{\mathrm{b}}$ the vector unit normal and $q_{\mathrm{ext}}$ the surface heat flux. \emph{In the EM equations} (\ref{eqn:potential})-(\ref{eqn:current}) and their BCs \eqref{eqn:fs_charge_conservation}-\eqref{eqn:ground} which enforce the conservation of charge; $\rho_{\mathrm{e}}$ is the electrical resistivity, $\psi$ an auxiliary potential and $J_{\mathrm{em}}$ the emitted current density. Three surface boundaries are distinguished;  $b_2$ is the free surface with the unit vector normal $\boldsymbol{\hat{n}}_{\mathrm{b}_2}$, $S_{\mathrm{i}}$ is any electrically insulated surface with the unit vector normal $\boldsymbol{\hat{n}}_{\mathrm{i}}$ and $S_{\mathrm{g}}$ is any grounded surface.

The volumetric heat sources in Eq.(\ref{eqn:energy}) include Joule heating and the Thomson effect, as well as possible sources from the incident plasma. Joule heating and thermoelectric effects, including Thomson heating, can be treated self-consistently with minor adjustments to the equations (\ref{eqn:energy}), (\ref{eqn:current}) and their boundary conditions, while plasma-induced volumetric heat sources have to be introduced as external input. In most fusion scenarios, the Joule and Thomson effects are negligible.

\section{Code structure}

Two versions of the MEMENTO code are available targeting three-dimensional (3D)  and two-dimensional (2D) problems. The following discussion applies to both dimensionalities, unless explicitly stated otherwise.

\begin{figure}[t]
  \centering
  \subfloat{\includegraphics[width=0.99\columnwidth]{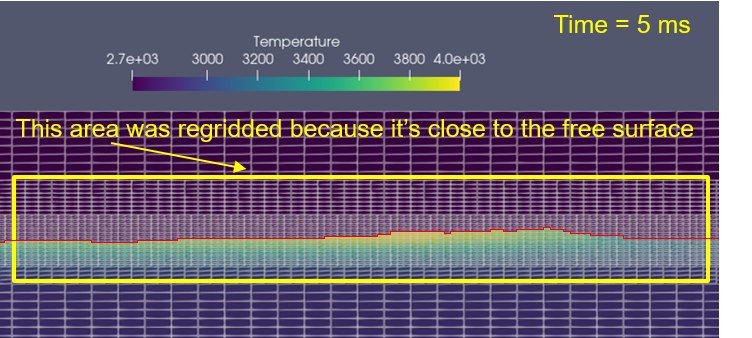} } \\[3pt]
  \subfloat{\includegraphics[width=0.97\columnwidth]{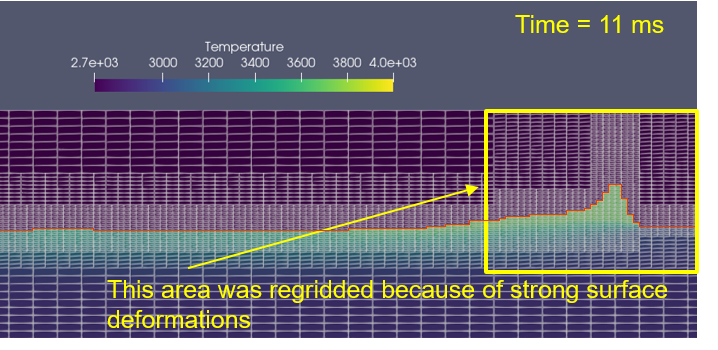} } %\\
  \caption{Two instants of a MEMENTO run that exemplify the different grid resolutions. The red line represents the free surface. The uniform dark-colored part above the free surface at the upper half of the gridded section is the background, while the lighter-colored part below the free surface is the simulated sample. Grey lines show the grid used.}
  \label{fig:regridding}
\end{figure}

MEMENTO relies on the AMReX framework~\cite{Amrex, Zhang2019_amrex} to generate and maintain an adaptive non-uniform grid. AMReX is a C++ software with a Fortran interface that provides functionality for solving partial differential equations with block-structured and adaptive mesh refinement algorithms. The non-uniform grid allows MEMENTO to perform highly resolved calculations near the free surface and to save computational time by introducing a coarser grid far from it. Each distinct grid resolution defines an AMReX level, with the free surface always corresponding to the finest level. The coarsest level is always designated as \enquote{level 0}. The terms \emph{higher} and \emph{finer} will be used interchangeably when referring to AMReX levels. The ratio between grid resolutions of two different levels is referred to as the \emph{refinement ratio}; it is defined in the input and has to be a power of two in all MEMENTO simulations. Fig.\ref{fig:regridding} shows an example of the cross-section of a simulated sample and the employed three-level grid. Owing to the adaptive nature of the mesh, the grid has changed between the 5 and 11\,ms instants due to the build-up of a strong surface deformation. To complement the non-uniform spatial mesh, sub-cycling in time can be enabled so that higher levels are advanced with a smaller time step. The ratio between the time steps of different levels is equal to their refinement ratio. In particular, with sub-cycling enabled, for every update of a coarser level, finer levels will undergo as many more updates as specified by the refinement ratio between them. For example, if there are two levels with a refinement ratio of 4, when advancing the coarse level in time once, the fine level is advanced four times with a 4 times smaller time step.

The code is split into several modules that assist the solution of the heat equation, SW equations and current propagation equations. Other modules of the MEMENTO code include the material library module, the initialization module as well as those for input and output. The material library is continuously updated and its latest version comprises the thermophysical properties of tungsten, beryllium, ATJ graphite, iridium, niobium, aluminum, tungsten heavy alloy and CuCrZr alloy. It is based on Refs.~\cite{tolias2017_1, TOLIAS2022_1, repo}.

The thermodynamic equation that connects the specific isobaric heat capacity with the specific enthalpy $d\phi=c_{\mathrm{p}}(T)dT$ is employed to generate $T(\phi)$ tabulations (with temperature increments specified in the input file). This aims to reduce the computational cost; whenever the code needs to translate an enthalpy to a temperature or vice versa it interpolates between the tabulated values instead of computing an integral. A mapping between enthalpy and temperature is necessary, since parts of the code are based on enthalpies (\emph{e.g.} the treatment of liquid-solid phase transitions), while other parts are based on temperatures (\emph{e.g.} the temperature-dependent thermophysical properties).

The heat and electrostatics solvers are parallelized utilizing AMReX's built-in capabilities. The computational domain is divided into boxes that are created automatically by the AMReX subroutines. The box size and the box number can be controlled by the input. In the current MEMENTO version, different boxes are advanced in parallel by using multiple OpenMP threads. At the edges of different boxes, ghost points are used to share information at the boundaries when required by the stencils.

\section{Fluid solver}

\begin{figure}
    \includegraphics[width=\columnwidth]{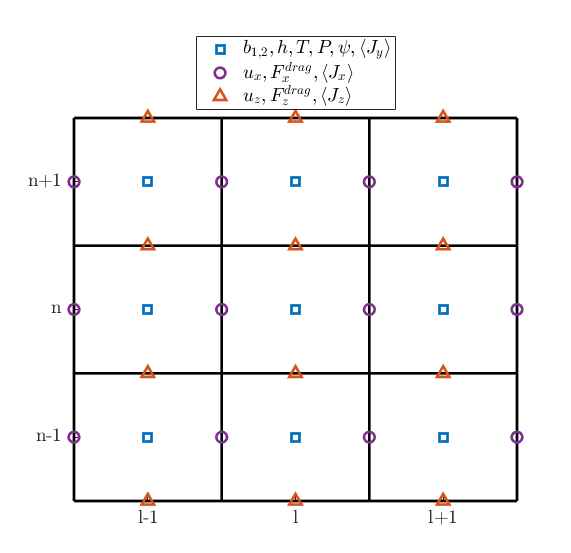}
    \caption{Sketch of the staggered grid employed in MEMENTO. It is noted that $\langle J_y \rangle$ is staggered on a plane perpendicular to this view.}
    \label{fig:MEMENTO_grid}
\end{figure}

The SW equations are solved on a 2D grid that is defined by the finest AMReX level and is aligned with the x-z plane. The grid employed is staggered as shown in Fig.\ref{fig:MEMENTO_grid}.

\subsection{Implementation}

The free surface position is obtained from an update of Eq.\eqref{eqn:column_height} discretized using the explicit upwind scheme
\begin{equation}
\label{eqn:column_height_simple}
    \begin{aligned}
        \frac{b_{2, l, n}^{i+1}-b_{2, l, n}^{i}}{\Delta t} &+ \frac{f_{l+1/2, n}^{u_{x}}-f_{l-1/2, n}^{u_{x}}}{\Delta x}
        \\
        &+ \frac{f_{l, n+1/2}^{u_{z}}-f_{l, n-1/2}^{u_{z}}}
        {\Delta z} = - \textcolor{Black}{\dot{x}_{\mathrm{vap}, l, n}^{i}}, \\
    \end{aligned}
\end{equation}
where
\begin{align}
    \nonumber
    &\begin{aligned}
        f_{l+1/2, n}^{u_{x}} &= \frac{u_{x, l+1/2, n}^{i}+\left|u_{x, l+1/2, n}^{i}\right|}{2} h_{l, n}^{i}
        \\
        &+ \frac{u_{x, l+1/2, n}^{i}-\left|u_{x, l+1/2, n}^{i}\right|}{2} h_{l+1, n}^{i},
    \end{aligned}
    \\
    \nonumber
    &\begin{aligned}
        f_{l, n+1/2}^{u_{z}} &= \frac{u_{z, l, n+1/2}^{i}+\left|u_{z, l, n+1/2}^{i}\right|}{2} h_{l, n}^{i}
        \\
        &+ \frac{u_{z, l, n+1/2}^{i}-\left|u_{z, l, n+1/2}^{i}\right|}{2} h_{l, n+1}^{i}.
    \end{aligned}
\end{align}
All but the first term in Eq.\eqref{eqn:column_height_simple} are directly dependent on the heat solver.

The depth-averaged velocity field $\boldsymbol{U}=\{u_x, u_z\}$ is also progressed with an explicit upwind scheme. Therefore, the $x$-component of Eq.(\ref{eqn:momentum}) is updated as
\begin{multline}
    \label{eqn:momentum_simple}
    \frac{u_{x, l+1/2, n}^{i+1}-u_{x, l+1/2, n}^{i}}{\Delta t}+f_{l+1/2, n}^{u_{x}, u_{x}}+f_{l+1/2, n}^{u_{z}, u_{x}} = g_x + \frac{1}{\textcolor{Black}{\rho_{\mathrm{m}}}}
    \\
    \Bigg[ \textcolor{Black}{ F_{x,l+1/2,n}^{\mathrm{J \times B}} }-\frac{3 \textcolor{Black}{\mu}}{\left(\text{max}\left( \left[ \tilde{h}_{l+1, n} + \tilde{h}_{l, n}\right]/2, h_{\mathrm{cap}} \right) \right)^{2}} u_{x, l+1/2, n}^{i+1}
    \\
    +L_{l+1/2, n}^{i} + M_{l+1/2,n} + \textcolor{Black}{F^{\mathrm{pres}}_{l+1/2,n}}  + F^{\mathrm{drag}}_{x,l+1/2,n} +\textcolor{Black}{F_{l+1/2,n}^{\mathrm{ST}}} \Bigg],
\end{multline}
where
\begin{align*}
    &\begin{aligned}
        L_{l+1/2, n}^{i} &= \textcolor{Black}{\mu} \Big(\frac{u_{x, l+3/2, n}^{i}-2 u_{x, l+1/2, n}^{i}+u_{x, l-1/2, n}^{i}}{\Delta x^{2}} +
        \\
        &\hspace{0.5cm}\frac{u_{x, l+1/2, n+1}^{i}-2 u_{x, l+1/2, n}^{i}+u_{x, l+1/2, n-1}^{i}}{\Delta z^{2}}\Big),
    \end{aligned}
    \\
    &\begin{aligned}
        M_{l+1/2,n} &= \frac{3}{2 \text{max} \left(\left[\tilde{h}_{l+1,n}+\tilde{h}_{l,n}\right]/2, h_{\mathrm{cap}}^{\mathrm{mg}} \right)}
        \\
        &\hspace{0.5cm} {\frac{\partial \sigma}{\partial T}} \frac{T_{l+1,n}-T_{l,n}}{\sqrt{\Delta x^2 + \left(b_{2,l+1,n}-b_{2,l,n} \right)^2}} ,
    \end{aligned}
    \\
    &\textcolor{Black}{F^{\mathrm{pres}}_{l+1/2,n}} = \frac{ \textcolor{Black}{P_{l+1,n}}-\textcolor{Black}{P_{l,n}}}{\Delta x},
    \\
    &\begin{aligned}
        \textcolor{Black}{F^{\mathrm{ST}}_{l+1/2,n}} &= \textcolor{Black}{\sigma_{l+1/2,n}} \frac{1}{\Delta x \Delta z^2}   \Big( b^{i+1}_{2,l+1,n+1} + b^{i+1}_{2,l+1,n-1} - 2b^{i+1}_{2,l+1,n}
        \\
        &- b^{i+1}_{2,l,n+1} - b^{i+1}_{2,l,n-1} + 2b^{i+1}_{2,l,n} \Big)
        \\
        &+ \sigma_{l+1/2,n} \frac{1}{\Delta x^3}\Big( -b^{i+1}_{2,l-1,n} + 3b^{i+1}_{2,l,n}
        \\
        &- 3b^{i+1}_{2,l+1,n} + b^{i+1}_{2,l+2,n} \Big),
    \end{aligned}
\end{align*}
\begin{align*}
    &\begin{aligned}
        f_{l+1/2, n}^{u_{x}, u_{x}} &=\frac{u_{x, l+1/2, n}^{i}+\left|u_{x, l+1/2, n}^{i}\right|}{2}\left[\frac{u_{x, l+1/2, n}^{i}-u_{x,l-1/2, n}^{i}}{\Delta x}\right]
        \\
        &+ \frac{u_{x, l+1/2, n}^{i}-\left|u_{x, l+1/2, n}^{i}\right|}{2}\left[\frac{u_{x,l+3/2, n}^{i}-u_{x,l+1/2, n}^{i}}{\Delta x}\right],
    \end{aligned}
    \\
    &\begin{aligned}
        f_{l+1/2, n}^{u_{z}, u_x} &=\frac{u_{z, l, n-1/2}^{i}+\left|u_{z, l, n-1/2}^{i}\right|}{2}\left[\frac{u_{x, l+1/2, n}^{i}-u_{x, l+1/2, n-1}^{i}}{\Delta z}\right]
        \\
        &+ \frac{u_{z, l, n-1/2}^{i}-\left|u_{z, l, n-1/2}^{i}\right|}{2}\left[\frac{u_{x, l+1/2, n+1}^{i}-u_{x, l+1/2, n}^{i}}{\Delta z}\right].
    \end{aligned}
\end{align*}
The parameters $h_{\mathrm{cap}}$ and $h_{\mathrm{cap}}^{\mathrm{mg}}$ are controlled by the input and serve as ad-hoc solutions to possible numerical issues by capping the viscous damping and the Marangoni drive to a minimum effective fluid height. The designation $\tilde{h}_{l,n-1/2}$ refers to the partially updated height obtained by solving for the position of the free surface in time step $i+1$ but not for the solid-liquid interface, i.e. $\tilde{h}_{l,n} = b_{2,l,n}^{i+1} - b_{1,l,n}^{i}$. It is noted that the explicit formulation of the surface tension $F^{\mathrm{ST}}$ introduces a stringent stability criterion for the time step that scales poorly with the tangential discretizations $\Delta x$ and $\Delta z$~\cite{Kondic2001}. Moreover, the Lorentz force, $\boldsymbol{F}^{\mathrm{J \times B}}$, is a variable that has to be passed from the electrostatics solver with the following interpolations employed
\begin{align*}
    \textcolor{Black}{F_{x,l+1/2,n}^{\mathrm{J \times B}}} &= \Big(\langle J_{y, l,n} \rangle + \langle J_{y, l+1,n} \rangle\Big) B_z/2
    \\
    & - \Big( \langle J_{z, l+1,n-1/2} \rangle + \langle J_{z, l+1,n+1/2} \rangle
    \\
    &\hspace{0.5cm}+ \langle J_{z, l,n-1/2} \rangle + \langle J_{z, l,n+1/2} \rangle \Big) B_y/4.
\end{align*}
Finally, it might be desirable to bypass the self-consistent evaluation of the replacement current density to save computational time~\cite{Thoren2018_1}. Then, the Lorentz force can instead be approximated as
\begin{equation*}
    \textcolor{Black}{F_{x,l+1/2,n}^{\mathrm{J \times B}}} = c \left( \textcolor{Black}{ J_{\mathrm{em},l+1, n} + J_{\mathrm{em},l, n} } \right) B_{z}/2,
\end{equation*}
where $c$ is controlled from the input file and varies between zero and unity. It also possible to pass from input a single depth-averaged current density value. The equation for the $z-$component is analogous to Eq.(\ref{eqn:momentum_simple}).

\subsection{Boundary conditions}

\begin{figure}
    \includegraphics[width=\columnwidth]{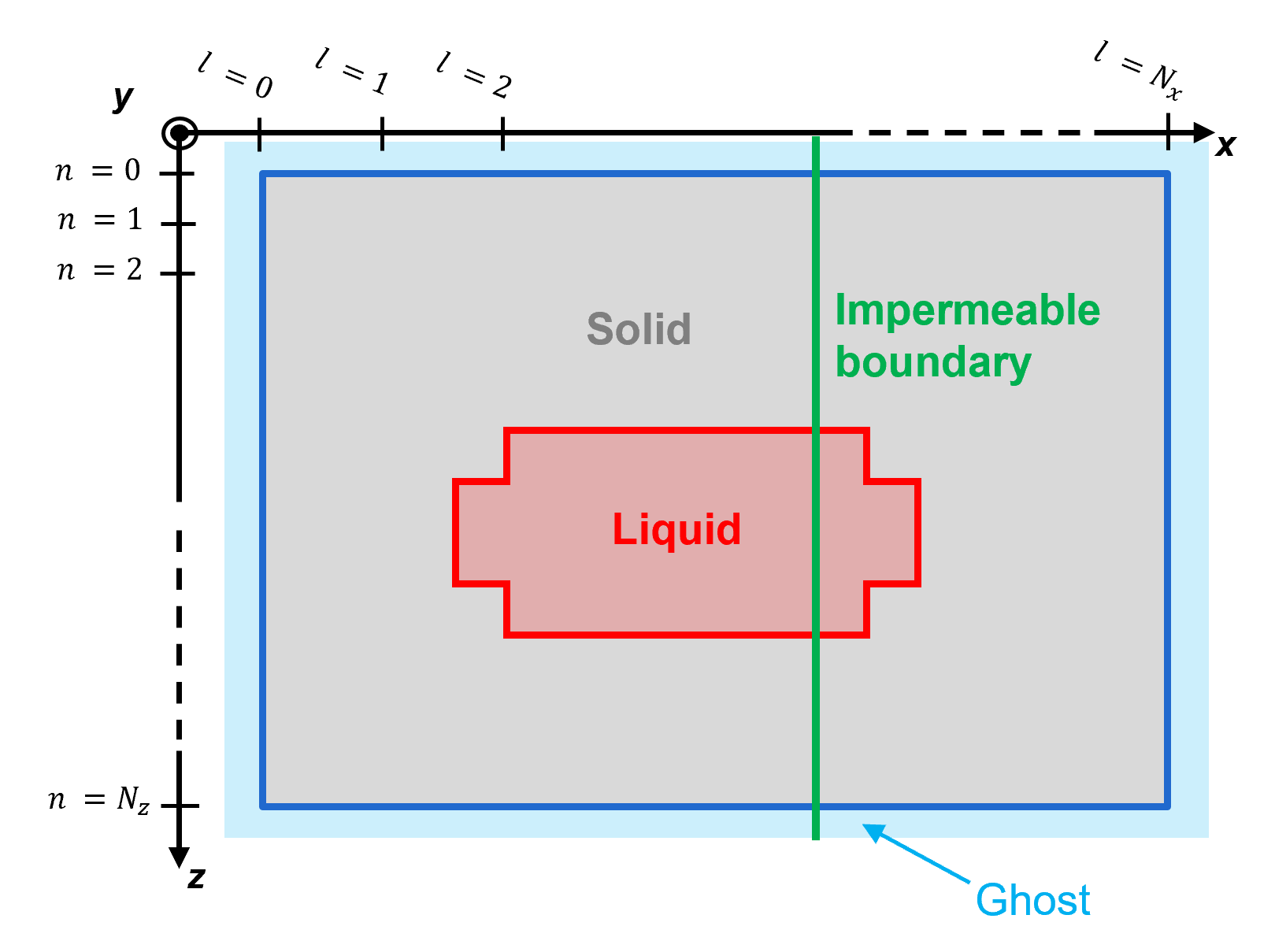}
    \caption{ Sketch of the computational domain used by the SW solver. The boundary of the computational domain is outlined in blue, the bound between the melt pool and the solid is outlined in red, and the impermeable boundary is drawn in green.}
    \label{fig:SW_boundary}
\end{figure}

Three types of boundaries are considered in the SW solver; the bounds of the computational domain, input-defined impermeable planes and the liquid-solid interface. The three types of boundaries are illustrated in Fig.\ref{fig:SW_boundary}.

For the computational domain boundary, a free outflow and no-inflow BC is enforced by default. For example, the BC at the $l=0$ edge is implemented with the statement: if $u_{x,1,n}<0$ then $u_{x,0,n}=u_{x,1,n}$, otherwise $u_{x,0,n}=0$. Similar statements are implemented on the $l=N_x$ edge for $u_{x,N_x,n}$ and on the $n=\{0, N_z\}$ edges for $u_{z,l,n}$. Alternatively, the user may opt to enforce impermeable BCs at the domain edges. Switching the default BC to impermeable boundaries at the domain edges is recommended in scenarios where the melt pools are created at the edge of the domain and particularly if the surface tension, $\boldsymbol{F}^{\mathrm{ST}}$, is enabled. In order to treat $F^{\mathrm{ST}}_{l+1/2,n}$ at the edge of the domain, due to the extended third derivative stencil, ghost points are used when necessary with $b_2$ set to be equal to the value of the nearest valid cell.

For impermeable planes inside the computational domain, the flux of material impeding through them is set to zero. For liquid-solid interfaces, a no-inflow of solid material is imposed. This occurs naturally when solving Eq.(\ref{eqn:column_height_simple}) even though the staggered grid might result in a non-zero velocity calculated by Eq.(\ref{eqn:momentum_simple}). To avoid issues when calculating the heat advection, an extra check is implemented in the SW solver so that any velocity that points outwards from a solid point is reset to zero.

\section{Heat transfer solver}

The heat transfer equation is solved on all AMReX levels. Two heat solvers are implemented in MEMENTO, one with an explicit formulation for the heat diffusion and one with an implicit formulation. Both solvers use a staggered grid, where, as shown in Fig.\ref{fig:MEMENTO_grid}, the temperature is defined in the cell center, while the velocity on the grid faces.

The velocity field employed by the heat solver, $\boldsymbol{U}_{\mathrm{h}}$, is a 3D vector, whereas the depth-averaged velocity $\boldsymbol U$ from the SW solver is a 2D vector. Reconstruction of $\boldsymbol{U}_{\mathrm{h}}$ from $\boldsymbol{U}= (u_x,u_z)$ is based on $\boldsymbol{U}_{\mathrm{h}} = (u_x,0,u_z)$ consistent with the SW assumption of $|u_x|, |u_z| \gg |u_y|$. For ease of notation, in the following, $\boldsymbol{U}_{\mathrm{h}}$ will be simply expressed as $\boldsymbol{U}$.

\subsection{Explicit solver}

The explicit heat solver is based on the enthalpy formulation of Eq.(\ref{eqn:energy}), \emph{i.e.},
\begin{equation}
    \label{eqn:enthalpy_eqn}
    \frac{\partial \phi}{\partial t} + \nabla \cdot \left( {\boldsymbol{U}}\phi - k \nabla T \right) = \phi \left( \nabla \cdot {\boldsymbol{U}} \right) + V.
\end{equation}
The velocity field is directly dependent on the SW solver. Note that the mass density has been incorporated in the definition of the enthalpy and that the depth-averaged velocity field $\boldsymbol{U}$ is not solenoidal. The enthalpy formulation allows a straightforward treatment of phase transitions~\cite{Hu1996}.

\subsubsection{Implementation} \label{section:explicit_heat}

The discretized scheme for Eq.(\ref{eqn:enthalpy_eqn}) reads as
\begin{multline}
    \label{eqn:heat_eqn_explicit}
    \frac{\phi_{l, m, n}^{i+1}-\phi_{l, m, n}^{i}}{\Delta t}+F_{l, m, n} =
    \\
    D_{l, m, n} + V_{l,m,n} + S_{l,m,n},
\end{multline}
where
\begin{align*}
    &\begin{aligned}
        F_{l, m, n} &= \frac{f_{l+1/2, m, n}^{u_{x}}-f_{l-1/2, m, n}^{u_{x}}}{\Delta x}
        \\
        &+\frac{f_{l, m+1/2, n}^{0}-f_{l, m-1/2, n}^{0}}{\Delta y}
        \\
        &+\frac{f_{l, m, n+1/2}^{u_{z}}-f_{l, m, n-1/2}^{u_{z}}}{\Delta z},
    \end{aligned}
    \\
    &\begin{aligned}
        D_{l, m, n}=\phi_{l, m, n}^{i}\Big(& \frac{u_{x, l+1/2, m, n}-u_{x, l-1/2, m n}}{\Delta x} +
        \\
        &\frac{u_{z, l, m, n+1/2}-u_{z, l, m, n-1/2}}{\Delta z}\Big),
    \end{aligned}
    \\
    \nonumber
    &\begin{aligned}
        f_{l+1/2, m, n}^{u_{x}} &= \frac{u_{x, l+1/2, m, n}^{*}+\left|u_{x, l+1/2, m, n}^{*}\right|}{2} \phi_{l, m, n}^{i}
        \\
        &+\frac{u_{x, l+1/2, m, n}^{*}-\left|u_{x, l+1/2, m, n}^{*}\right|}{2} \phi_{l+1, m, n}^{i} \\
        & - k_{l+1/2, m, n}^{i} \frac{T_{l+1, m, n}^{i}-T_{l, m, n}^{i}}{\Delta x},
        \end{aligned}
    \\
    \nonumber
    &f_{l, m+1/2, n}^{0}=-k_{l, m+1 / 2, n}^{i} \frac{T_{l, m+1, n}^{i}-T_{l, m, n}^{i}}{\Delta y},
    \\
    \nonumber
    &\begin{aligned}
        f_{l, m, n+1/2}^{u_{z}} &= \frac{u_{z, l, m, n+1/2}^{*}+\left|u_{z, l, m, n+1/2}^{*}\right|}{2} \phi_{l, m, n}^{i}
        \\
        & +\frac{u_{z, l, m, n+1/2}^{*}-\left|u_{z, l, m, n+1/2}^{*}\right|}{2} \phi_{l, m, n+1}^{i}
        \\
        &- k_{l, m, n+1 / 2}^{i} \frac{T_{l, m, n+1}^{i}-T_{l, m, n}^{i}}{\Delta z}.
    \end{aligned}
\end{align*}

The volumetric source term $V_{l,m,n}$ is an input to the code, as is $S_{l,m,n}$ which stems from the surface heat loads and is discussed in section \ref{section:explicit_heat_BC}. The symbol $k_{l+1/2,m,n}^i$ represents the temperature-dependent thermal conductivity at the interface between the cells $(l+1, m, n)$ and $(l, m, n)$. The thermal conductivity at that interface is evaluated at the temperature $\left( T_{l,m,n}^i + T_{l-1,m,n}^i \right)/2$ with equivalent assumptions being made for $k_{l,m-1/2,n}^i$ and $k_{l,m,n-1/2}^i$. The time step indicator $^*$ can take two values. It is equal to $i+1$ when subcycling is disabled or the update is performed at the finest AMReX level. Alternatively, it is equal to $i$ when subcycling is enabled and the update is performed in levels lower than the highest one.

After the update described by Eq.(\ref{eqn:heat_eqn_explicit}), the temperature is obtained from the enthalpy by making use of the tabulated relation $T(\phi)$. This is necessary in order to compute all the material properties in the following time step. Finally, the new position of the melt bottom $b_1$ is stored.

\subsubsection{Boundary conditions}\label{section:explicit_heat_BC}

The boundary heat fluxes are converted to volumetric heat sources, $S_{l,m,n}$, by appropriately scaling the incoming flux, \emph{e.g.} by dividing with $1/\Delta y$ if the flux is applied on the $xz$-aligned face of a cell. The converted fluxes are applied to the first cell center that is adjacent to the respective interface. This is done because there are boundaries that do not coincide with the borders of the computational domain where the AMReX boundary conditions are applied. Conduction and advection across the free surface and the cooling pipe are set to zero. Furthermore, there is no advection at the liquid-solid interface. Finally, the explicit solver allows the use of a prescribed temperature instead of prescribed heat flux at the free surface as well as a prescribed temperature at the lowest xz-plane of the domain.

The heat flux parallel to the magnetic field lines is provided to the code as input. For the free-surface, the parallel heat flux is scaled based on an input-defined property, either an angle or the unit vector of the heat flux. When the unit vector is employed, its scalar product with unit vectors that are locally normal to the free surface define the scaling coefficient. The surface normals are found as $\hat{\boldsymbol{n}}_{\mathrm{b}_2} = \nabla b_2/ |\nabla b_2|$ where in the discretized form a second order central scheme is used for non-domain-boundary points and forward or backward first order schemes are utilized otherwise. In any surfaces, the scaling of the parallel flux can only be carried out with the input-defined angle.

\subsection{Implicit solver}

\subsubsection{Implementation}

The implicit solver follows a three-step procedure. First, a partial update is obtained by solving only the conductive part of Eq.(\ref{eqn:energy}), \emph{i.e.},
\begin{equation}
    \label{eqn:temperature_conduction}
    c_{\mathrm{p}}\rho_{\mathrm{m}}\frac{\partial{T}}{\partial{t}}=\nabla\cdot\left(k\nabla{T}\right)+V,
\end{equation}
using an implicit Euler method that employs one of the linear solvers implemented in the AMReX package~\cite{Amrex, Zhang2019_amrex}. The user of MEMENTO can choose the preferred solver via an input parameter. The AMReX solvers require a linear system resulting from the canonical form
\begin{equation*}
    \left( A \alpha - B \nabla \cdot \beta \nabla \right) \chi = f.
\end{equation*}
In MEMENTO, it is set that
\begin{align*}
    A &= 1, \\
    B &= \Delta \text{t}, \\
    \alpha_{l,m,n} &= \rho^i_{\mathrm{m}-l,m,n} c^i_{\mathrm{p}-{l,m,n}}, \\
    \beta^x_{l+1/2,m,n} &= k^i_{l+1/2,m,n}, \\
    \chi_{l,m,n} &= T^{i+1}_{l,m,n}, \\
    f_{l,m,n} &= S_{l,m,n} \Delta \text{t} + \alpha_{l,m,n}T_{l,m,n}^i + V_{l,m,n} \Delta \text{t}.
\end{align*}
The term $S_{l,m,n}$ has been discussed in section \ref{section:explicit_heat_BC}. After the update described by Eq.\eqref{eqn:temperature_conduction}, the heat integration method~\cite{Hu1996} is used to compute the enthalpy of all points undergoing a phase transition. The temperature of cells undergoing a phase transition is kept constant at the melting temperature $T_{\mathrm{m}}$ until the enthalpy of fusion has been accumulated. For the remaining points, the enthalpy can be directly obtained by inverting the tabulated relation $T(\phi)$. Finally, the update is completed by advancing the advective part of Eq.(\ref{eqn:energy}), which is written in terms of enthalpy. The scheme used is analogous to the one employed in the explicit heat solver and reads out as
\begin{equation}
    \label{eqn:enthalpy_advection}
    \frac{\phi_{l, m, n}^{i+1}-\phi_{l, m, n}^{i}}{\Delta t}+\bar{F}_{l, m, n}=D_{l, m, n},
\end{equation}
where
\begin{align}
    \nonumber
    &\bar{F}_{l, m, n}=\frac{\bar{f}_{l+1/2, m, n}^{u_{x}}-\bar{f}_{l-1/2, m, n}^{u_{x}}}{\Delta x}+\frac{\bar{f}_{l, m, n+1/2}^{u_{z}}-\bar{f}_{l, m, n-1/2}^{u_{z}}}{\Delta z}, \\
    \nonumber
    &\begin{aligned}
        \bar{f}_{l+1/2, m, n}^{u_{x}} &= \frac{u_{x, l+1/2, m, n}^{*}+\left|u_{x, l+1/2, m, n}^{*}\right|}{2} \phi_{l, m, n}^{i}
        \\
        &+\frac{u_{x, l+1/2, m, n}^{*}-\left|u_{x, l+1/2, m, n}^{*}\right|}{2} \phi_{l+1, m, n}^{i},
    \end{aligned}
    \\
    \nonumber
    &\begin{aligned}
        \bar{f}_{l, m, n+1/2}^{u_{z}} &= \frac{u_{z, l, m, n+1/2}^{*}+\left|u_{z, l, m, n+1/2}^{*}\right|}{2} \phi_{l, m, n}^{i}
        \\
        &+\frac{u_{z, l, m, n+1/2}^{*}-\left|u_{z, l, m, n+1/2}^{*}\right|}{2} \phi_{l, m, n+1}^{i}.
    \end{aligned}
\end{align}
The time step indicator $^*$ has the same meaning as the one discussed in the explicit heat solver, see section \ref{section:explicit_heat}.

\subsubsection{Boundary conditions} \label{section:implicit_heat_BC}

The boundary conditions of the implicit solver are identical to the ones described in section \ref{section:explicit_heat_BC}, with the exception that the implicit solver cannot work with a prescribed temperature at any surface.

\section{Electrostatics solver}

The electrostatics solver computes the replacement current density via the auxiliary potential by solving Eq.(\ref{eqn:potential}) on all AMReX levels and Eq.(\ref{eqn:current}) only on the highest level and within the melt pool. In a given time step, the solver reuses the same grid that the heat solver utilized. The auxiliary potential and current density, are defined on a staggered grid; the former at the cell centers while the latter at the cell faces, as shown in Fig.\ref{fig:MEMENTO_grid}.

\subsection{Implementation}

The implementation relies on linear solvers readily available through the AMReX interface to obtain a solution to Eq.\eqref{eqn:potential}. As in the implicit heat solver, the AMReX interface to the linear solver requires an equation of the form
\begin{equation*}
    \left( A \alpha - B \nabla \cdot \beta \nabla \right) \chi = f.
\end{equation*}
To translate this in the discretized form of Eq.\eqref{eqn:potential}, it is set that
\begin{align*}
    A &= 1, \\
    B &= -1, \\
    \alpha &= 0, \\
    \beta^x_{l+1/2,m,n} &= 1/\rho_{\mathrm{e}-l+1/2,m,n}, \\
    \chi_{l,m,n} &= \psi_{l,m,n}, \\
    {f}_{l,m,n} &=  \begin{array}{ll}
         &  P_\mathrm{d} - G_\mathrm{d} + J_{\mathrm{em}-l, n}/{\Delta y}, \text{ if }m=m^y_{\mathrm{b}_2}  \\
         & {P_\mathrm{d}}, \text{ otherwise}
    \end{array}
\end{align*}
with $m^{y,l,n}_{\mathrm{b}_2}$ the maximum integer for which the $(l,m^{y,l,n}_{\mathrm{b}_2},n)$ cell falls below the free surface and
\begin{align*}
    {P}_{\mathrm{d}} = &- \frac{B_y}{\Delta x} \left[ \frac{u^{i+1}_{z,l+1/2,m,n}}{\rho_{\mathrm{e}-l+1/2,m,n}}  - \frac{u^{i+1}_{z,l-1/2,m,n}}{\rho_{\mathrm{e}-l-1/2,m,n}}  \right]
    \\
    &+ \frac{B_y}{\Delta z} \left[ \frac{u^{i+1}_{x,l,m,n+1/2}}{\rho_{\mathrm{e}-l,m,n+1/2}}  - \frac{u^{i+1}_{x,l,m,n-1/2}}{\rho_{\mathrm{e}-l,m,n-1/2}}  \right]
    \\
    &+\frac{1}{2 \Delta y} \left(B_x u^{i+1}_{z,l,m,n} - B_z {u^{i+1}_{x,l,m,n}}\right)
    \\
    &\left[\frac{1}{\rho_{\mathrm{e}-l+1,m,n}}  - \frac{1}{\rho_{\mathrm{e}-l-1,m,n}}  \right] + \Gamma_d
\end{align*}
and
\begin{align*}
    {G}_\mathrm{d} &= \frac{1}{\rho_\mathrm{e-l,m,n}} \Big[- \hat{n}_{x,l,n} u^{i+1}_{z,l,m,n} B_y \frac{1}{\Delta x}
    \\
    &+ \hat{n}_{y,l,n} \left(u^{i+1}_{z,l,m,n} B_x - u^{i+1}_{x,l,m,n} B_z \right)\frac{1}{\Delta y}
    \\
    &+\hat{n}_{z,l,n} u^{i+1}_{x,l,m,n} B_y \frac{1}{\Delta z} \Big],
    \\
    \Gamma_d &= \frac{B_y}{\rho_{\mathrm{e}-l-1,m,n}} \Big[p_{x,l} \frac{0.5}{\Delta x}(u^{i+1}_{z,l,m,n+1} + u^{i+1}_{z,l,m,n})
    \\
    &-p_{z,n} \frac{0.5}{\Delta z}(u^{i+1}_{x,l+1,m,n} + u^{i+1}_{x,l,m,n}) \Big].
\end{align*}
The scalars $p_{x,l}$ and $p_{z,n}$ are $-1$ at $l=0$ and $n=0$ respectively, they are $1$ at $l=l_{max}$ and $n=n_{max}$ and 0 otherwise. The surface normal, as discussed in section \ref{section:explicit_heat_BC}, is found as $\hat{\boldsymbol{n}}_{\mathrm{b}_2} = \nabla b_2/|\nabla b_2|$ using a second order central scheme for points not adjacent to the domain boundary and a first order forward or backward scheme otherwise.

After the auxiliary potential is found, the current density due to potential gradients is computed as
\begin{equation}
    \label{eqn:current_implementation}
    J^{\mathrm(p)}_{x,l-1/2,m,n} = - \frac{1}{\rho_{\mathrm{e}-l-1/2,m,n}} \frac{\psi_{l,m,n}-\psi_{l-1,m,n}}{\Delta x}
\end{equation}
with $\rho_{\mathrm{e}-l-1/2,m,n}$ the electrical resistivity at the interface between the cells $(l-1,m,n)$ and $(l,m,n)$ calculated at the temperature $( T_{l,m,n}^i + T_{l-1,m,n}^i)/2$. The expressions for $J^{\mathrm(p)}_{y,l,m-1/2,n}$ and $J^{\mathrm(p)}_{z,l,m,n-1/2}$ are completely analogous to Eq.(\ref{eqn:current_implementation}). Finally, the depth-averaged current values within each melt column, defined by Eq.\eqref{eqn:current_analytical_average}, are computed as
\begin{multline}
    \label{eqn:current_average}
    \langle J \rangle_{x, l-1/2,n} =  \frac{1}{m^{x,l-1/2,n}_{\mathrm{b}_2}-m^{x,l-1/2,n}_{\mathrm{b}_1}+\left| m_{\mathrm{b}_2}^{y,l,n} - m_{\mathrm{b}_2}^{y,l-1,n}\right|}
    \\
    \hspace{0.2cm} \Bigg(\hat{\boldsymbol x} \cdot \hat{\boldsymbol{n}}_{\mathrm{fs}} \frac{-J_{\mathrm{em}-l-1,n} - J_{\mathrm{em}-l,n} }{2}\left| m_{\mathrm{b}_2}^{y,l,n} - m_{\mathrm{b}_2}^{y,l-1,n}\right|
    \\
    \hspace{0.0cm}+\sum_{m=m^{x,l-1/2,n}_{\mathrm{b}_1}}^{m^{x,l-1/2,n}_{\mathrm{b}_2}} J^{\mathrm(p)}_{x, l-1/2,m,n} \Bigg)\, -\frac{B_y}{\rho_{\mathrm{e}-l-1/2,m,n}}u_{z,l-1/2,m,n}.
\end{multline}
Here $m^{x,l-1/2,n}_{\mathrm{b}_1}$ is the minimum integer number for which either of the two cells $(l-1, m^{x,l-1/2,n}_{\mathrm{b}_1}, n)$ or $(l, m^{x,l-1/2,n}_{\mathrm{b}_1}, n)$ are above the re-solidification front, $m^{x,l-1/2,n}_{\mathrm{b}_2}$ is the maximum integer number for which both cells $(l-1,m^{x,l-1/2,n}_{\mathrm{b}_2}, n)$ and $(l, m^{x,l-1/2,n}_{\mathrm{b}_2}, n)$ fall below the position of the free surface.

\subsection{Boundary conditions}

The total current density at the free surface is given by the escaping emitted current which is typically equal to the thermionic current density and the $\boldsymbol{U} \times \boldsymbol{B}$ contribution. The surface current density is converted to a volumetric current source that is applied on the center of the first cell under the free surface, as is the case with the surface heat flux discussed in section \ref{section:explicit_heat_BC}. If the simulation geometry includes a cooling pipe, then the electrical conductivity at the interface of the cooling pipe is suppressed. In these surfaces, to avoid issues with floating cells, the conductivity is set to $10^{-16}$ instead of 0. The boundary conditions for the insulated and grounded surfaces, Eq.\eqref{eqn:insulation}-\eqref{eqn:ground}, imposed on the computational domain boundaries rely on native AMReX implementations for homogeneous conditions as the RHS of Eq.\eqref{eqn:insulation} has been added already through $\Gamma_d$.

\section{MEMENTO simulations}\label{section:simulation}

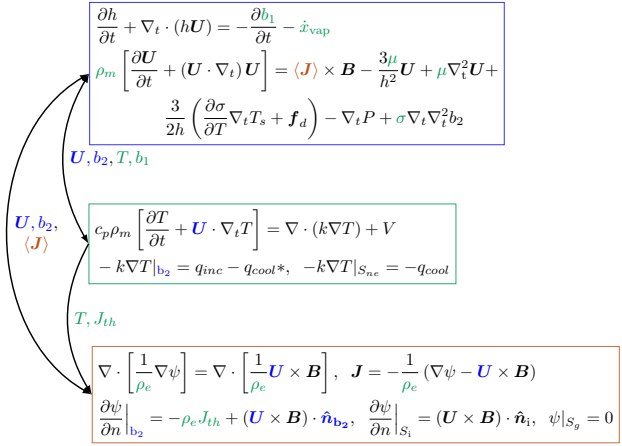
\begin{figure}
    \centering
    \resizebox{8.5cm}{!}{% \documentclass[14p]{article}
% \usepackage[dvipsnames]{xcolor}
% \usepackage[utf8]{inputenc}
% \usepackage{tikz}
% \usepackage{amsmath}
% \usetikzlibrary{shapes.geometric, arrows}
% \usetikzlibrary{automata, positioning}
% \usepackage{multirow}

% \def\mathcolor#1#{\@mathcolor{#1}}
% \def\@mathcolor#1#2#3{%
%   \protect\leavevmode
%   \begingroup
%     \color#1{#2}#3%
%   \endgroup
% }
% \tikzstyle{arrowc} = [thick,-->,>=stealth]

% \begin{document}
    \begin{tikzpicture}[tips=proper]

        % Setup the style for the states
        \tikzset{heat_node/.style={state, 
                                    fill=gray!0!white,
                                    draw=ForestGreen!80,
                                    rectangle}}
        \tikzset{SW_node/.style={state, 
                                    fill=gray!0!white,
                                    draw=blue!80,
                                    rectangle}}
        \tikzset{current_node/.style={state, 
                                    fill=gray!0!white,
                                    draw=Bittersweet!80,
                                    rectangle}}
        % Draw the states
        \node[SW_node] at (-6, 0)      (SW)     {
        $
            \begin{aligned}
                &\frac{\partial h}{\partial t}+\nabla_{t} \cdot(h \boldsymbol{U})=-\frac{\partial \mathcolor{ForestGreen}{b_{1}}}{\partial t}-\mathcolor{ForestGreen}{\dot{x}_{\mathrm{vap}}} 
                \\
                &\mathcolor{ForestGreen}{\rho_{m}}\left[\frac{\partial \boldsymbol{U}}{\partial t}+\left(\boldsymbol{U} \cdot \nabla_{t}\right) \boldsymbol{U}\right] = \mathcolor{Bittersweet}{\langle\boldsymbol{J}\rangle} \times \boldsymbol{B}-\frac{3 \mathcolor{ForestGreen}\mu}{h^{2}} \boldsymbol{U}+\mathcolor{ForestGreen}\mu \nabla_{\mathrm{t}}^{2} \boldsymbol{U} + \\ &\hspace{1.4cm} \frac{3}{2h}\left(\frac{\partial \sigma}{\partial T} \nabla_t T_s +\boldsymbol{f}_d \right) - \nabla_t P + \mathcolor{ForestGreen}\sigma \nabla_t \nabla_t^2 b_2
            \end{aligned}
        $
        };
        
        \node[heat_node] at (-6.5, -3.5)      (HT)     {
            $ 
        \begin{aligned}
            &c_{p} \rho_{m}\left[\frac{\partial T}{\partial t}+\mathcolor{blue}{\boldsymbol{U}} \cdot \nabla_{t} T\right]=\nabla \cdot(k \nabla T) + V
            \\
            &-k\nabla T|_{\mathcolor{blue}{{\mathrm{b}_2}}} = q_{inc}-q_{cool}*, \hspace{0.2cm} -k\nabla T|_{S_{ne}} = -q_{cool}
        \end{aligned}
        $
        };
        
        \node[current_node] at (-4.8, -6.6)      (CS)     {
            $
            \begin{aligned}
                &\nabla \cdot \left[ \frac{1}{\mathcolor{ForestGreen}{\rho_e}} \nabla \psi \right] = \nabla \cdot \left[ \frac{1}{\mathcolor{ForestGreen}{\rho_e}} \mathcolor{blue}{\boldsymbol{U}} \times \boldsymbol{B} \right], \hspace{0.2cm} \boldsymbol{J} = - \frac{1}{\mathcolor{ForestGreen}{\rho_e}} \left( \nabla \psi - \boldsymbol{\mathcolor{blue}U} \times \boldsymbol{B} \right)
                \\
                &\frac{\partial \psi}{\partial n}\Bigr|_{\mathcolor{blue}{{\mathrm{b}_2}}} = -\mathcolor{ForestGreen}{\rho_e J_{th}} + \left( \mathcolor{blue}{\boldsymbol{U}} \times \boldsymbol{B} \right) \cdot \mathcolor{blue}{\boldsymbol{\hat{n}_{\mathrm{b}_2}}},  \hspace{0.2cm} \frac{\partial \psi}{\partial n}\Bigr|_{S_{\mathrm{i}}} = \left( {\boldsymbol{U}} \times \boldsymbol{B} \right) \cdot {\boldsymbol{\hat{n}}}_{\mathrm{i}}, \hspace{0.2cm}  \psi|_{S_g} = 0 
            \end{aligned}
            $
        };

        % Connect the states with arrows
        \draw[<->,
              auto=right,
              line width=0.3mm,
              >=latex,
              draw=black,
              fill=black]
            (SW.west)     edge[bend right=30, auto=left] node {$\mathcolor{blue}{{\boldsymbol U}, b_2}, \mathcolor{ForestGreen}{T, b_1}$} (HT.west);

        \draw[<->,
              auto=right,
              line width=0.3mm,
              >=latex,
              draw=black,
              fill=black]
            (SW.west)     edge[bend right=60, auto=left] node[align=center] {$\mathcolor{blue}{{\boldsymbol U}, b_2}$, \\ $\mathcolor{Bittersweet}{\langle \boldsymbol{J} \rangle}$} (CS.west);

        \draw[->,
              auto=right,
              line width=0.3mm,
              >=latex,
              draw=black,
              fill=black]
            (HT.west)     edge[bend right=30, auto=left] node {$\mathcolor{ForestGreen}{T, J_{th}}$} (CS.west);
            % (HT)     edge[bend right=20, auto=left] node {$\mathcolor{ForestGreen}{T, b_1}$} (SW)
            % (CS)     edge[bend right=30, auto=right] node {$\mathcolor{Bittersweet}{\boldsymbol J}$} (SW)
            % (CS.west)     edge[bend left=60, auto=left] node {} (SW.west);
            % (P1)     edge[bend left=40, auto=left] node {} (P3)
            
            % (P2)     edge[bend left=20, auto=left]  node {} (P1)
            
            % (P3)     edge[bend left=40]            node {} (P1);
    \end{tikzpicture}
% \end{document}
}
    \caption{The coupling between the equations solved by the MEMENTO code. The shallow water, heat transfer and electrostatics modules are boxed in blue, green and red, respectively. Note that the wetted surfaces include but are not limited to $b_2$.}
    \label{fig:MEMENTO_coupling}
\end{figure}

In a MEMENTO simulation the shallow water, heat and electrostatics solvers are called consecutively exchanging information. The coupling between the solvers is illustrated in Fig.\ref{fig:MEMENTO_coupling}. A more detailed description is presented in \ref{sec:appendix} which includes a detailed flowchart.

Possible simulated geometries are categorized in two groups; slab and complex. In slab geometries the initial free surface lies at a constant height, $b_2(x,z,t=0) = y_0$, while in complex geometries the free surface is a function of position, $b_2(x,z,t=0) = y(x,z)$. Examples of complex geometries include, but are not limited to, free surfaces that are initially sloped, curved or step-like. The free surface is wetted by default, whereas the wetting of other surface boundaries is controlled by the input. Cooling is possible from all surfaces, but mass-flux due to vaporization is only defined for the free surface. The sample can include a heat sink in the solid which emulates a cooling pipe, as the one present in ITER monoblocks. Finally, it is possible to include regions where the thermal and electrical conductivities are set to zero, to emulate gaps in the domain.

In all simulations in which deformation is expected to take place, the computational domain has to be larger than the pristine sample and hence there are cells above the free surface that are not part of the simulated material. The gap between the top edge of the computational domain and the sample will be referred to as \textit{background}, see the dark purple colored cells in Fig.\ref{fig:regridding}.

As the domain evolves during a simulation, an integer field is used to flag and keep track of different regions. The background is marked with the value 0, the cooling pipe is assigned to -1, gaps are set to -2, while the solid, melting zone and liquid have the values 1, 2 and 3, respectively.

\section{Re-gridding, re-evaluation and synchronization}\label{section:regrid_and_revaluation}

In between advancing Eqs.\eqref{eqn:column_height_simple}-\eqref{eqn:current_average}, MEMENTO performs the following procedures; re-gridding, re-evaluation of the temperature due to deformation and synchronization of the different AMReX levels.

\emph{Re-gridding} refers to the modification of the grid used by the heat and electrostatics solvers to comply with the deformation of the free surface. The goal is to keep the finely resolved region to the minimum size required at each time-instant in order to reduce the computational cost. In the current version of MEMENTO, there are three defined criteria that can trigger re-gridding to a higher or lower level. Those are the distance of a point from the free surface, the surface deformation rate $\partial b_2 /\partial{t}$ and the material phase at a certain point. The values of the distance from the free surface and the deformation rate that trigger re-gridding are defined in the input. Note that the surface deformation rate is a function of $(x,z)$ and not $y$. Hence, if the value $(\partial b_2 / \partial t)|_{(x_0, z_0)}$ triggers re-gridding then all the $(x_0, y, z_0)$ points for which $y>b_2(x_0, z_0)$ are re-gridded. The third criterion sets all the points in the liquid phase at the highest level, thus the re-gridding is triggered as the melt front $b_1$ propagates. By default these criteria are checked at every time step while different frequencies can be chosen via input. The default option is to check for re-gridding at every time step. In Fig.\ref{fig:regridding}, at 5\,ms re-gridding has been triggered only due to the distance from the free surface while at 11\,ms also the surface deformation rate triggers re-gridding.

\emph{Re-evaluation} of the domain concerns the initialization of the temperature of the cells that were part of the background before the SW advance but fell below the free surface $b_2$ after the update, and vice-versa. The procedure is the following; material points that are added on top of the liquid are assigned the temperature of the pre-existing liquid, material points that are added on top of the solid are assigned the upwind temperature and material points that are removed are assigned the background temperature of zero Kelvin. Note that in the 3D MEMENTO version, a cell might have an upwind neighbor in the $x$-direction and one in the $z$-direction. Thus, the upwind temperature becomes ambiguous. In such cases, the used upwind temperature corresponds to its neighbour in the direction of which most mass flowed from. Domain re-evaluation is one of the first steps each time the heat solver is called.

\emph{Level synchronization} addresses a mismatch of heat fluxes at the boundary of two AMReX levels. The strategy adopted to advance the heat and electrostatics solvers is to first update the coarsest level, ignoring any finer level and then to progressively advance the finer levels using the coarser ones only to supply the boundary conditions at the fine-coarse level interface. At the end of the procedure outlined above, the fluxes from the advancement of the finer grid do not necessarily exactly match the underlying fluxes from the coarser grid faces. Thus, in order to avoid error accumulation over time, this mismatch needs to be corrected by synchronising the solution. The synchronization is performed by overwriting the values at the coarse cells to be the average of the values of the overlying fine cells. More details on how AMReX performs level synchronization can be found in the AMReX manual~\cite{Amrex}. Synchronization is necessary in the heat solver, but not in the electrostatics solver. Within one time step, the synchronization changes the solution merely on the coarser levels and not on the finest one. In the heat solver, as the solution of the next time step depends on the solution of the previous time step, non-synchronized levels would introduce errors at the highest level through the BC. In the electrostatics solver, however, the solution in each time step does not depend on the current density at the previous time step. Thus, non-synchronised levels can not alter the solution at the finest level. Since the electrostatics solver is used to find the current distribution in the melt pool, which always lies on the highest level, there is no need to spend computational power to synchronise the levels of the electrostatics solver.

\section{Solver benchmarking and comparison with experiments}\label{section:MEMENTObenchmarking}

Before illustrating the MEMENTO capabilities through the simulation of a tokamak experiment, the code benchmarking detailed in Ref.\cite{Paschalidis2023} will be summarized. The fluid shallow water solver has been compared with the augmented solver of the GeoClaw software~\cite{clawpack, LeVeque2011, Berger2011, Mandli2016}. The accuracy of the explicit heat solver has been checked against analytical solutions of a two-region Neumann problem~\cite{Bejan2003}. The implicit heat solver has been benchmarked utilizing the explicit heat solver as reference code. Furthermore, the MEMENTO implementation of a cooling pipe, based on simplifications suggested in Ref.\cite{Escourbiac2019}, has been compared to the results of ANSYS simulations~\cite{Panayotis2017} and the results of simulations with a dedicated solver~\cite{Van_den_Kerkhof2021}. Finally, the electrostatics solver has been validated against simulations with the COMSOL Multiphysics® software~\cite{Paschalidis2023}. The numerical inaccuracies were always found to be negligible compared to typical uncertainties of the MEMENTO input, primarily the experimental heat fluxes which often feature uncertainties in excess of 20$\%$~\cite{Ratynskaia2020,Thoren2021,Thoren2018,Ratynskaia2022_1}.

The validation of the physics model against multiple dedicated EUROfusion experiments has been presented in Refs.\cite{Ratynskaia2020,Ratynskaia2021,Thoren2021,Thoren2018,Ratynskaia2022_1}. While initial simulations had to rely mostly on a comparison with the final deformation profile, the latest experiments were designed in a way that provided additional novel experimental constraints such as the in-situ detection of the melt motion onset or the simultaneous exposure of samples with drastically different thermophysical properties. The model has been demonstrated to reliably reproduce experiments in several tokamaks (ASDEX-Upgrade, JET, WEST) that featured different PFC compositions (Be, W, Ir, Nb) and exposure geometries (leading edge, sloped, steps, gaps), that concerned various types of plasma scenarios (ELMing H-mode, disruptions) implying different sources of the replacement current (thermionic emission, halo current), that realized different electrical connections (grounded, insulated) and cooling mechanisms (active, passive)~\cite{Ratynskaia2020, Ratynskaia2021, Thoren2021,Thoren2018,Ratynskaia2022_1,Ratynskaia2023}.

\section{Example simulation of a tokamak experiment}\label{section:MEMENTOscenario}

Given the above, the simulation of a particular recent tokamak experiment was selected in order to demonstrate the capabilities of MEMENTO. This experiment, featuring several momentum sources, a non-trivial initial free surface, a non-slab sample geometry and a particular plasma wetting pattern, exemplifies a complex scenario which MEMENTO has been designed to model.

\subsection{Experimental scenario}

The exposure of an iridium (Ir) sample to the ASDEX Upgrade (AUG) tokamak discharges will be modelled. The main physics aspects have been reported in Ref.~\cite{Ratynskaia2022_1}. In what follows, the focus lies on numerical considerations concerning the simulation set-up and post-processing.

\begin{figure}
    \centering
    \includegraphics[width=\columnwidth]{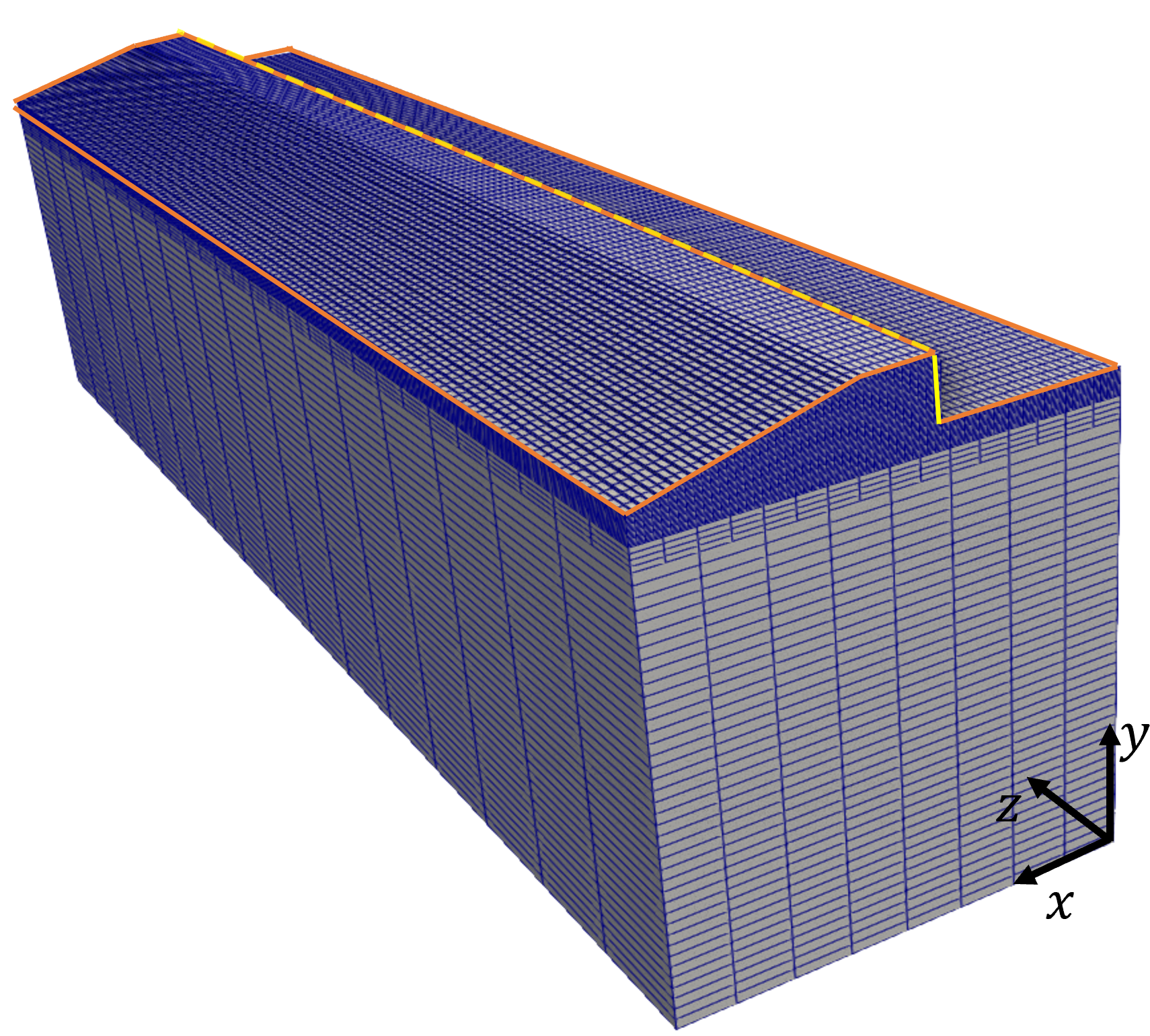}
    \caption{The simulated domain and the grids employed in modelling of the Ir ELM-induced melting experiment in the AUG tokamak. The plasma-wetted surface is outlined by an orange contour. Note that the surface outlined in yellow is not part of the free surface.}
    \label{fig:ir_domain}
\end{figure}

Recall that the incident heat flux constitutes an experimental input to MEMENTO. The sample was exposed during ELMing H-mode discharges that are characterized by an intermittent heat flux, where quiet periods of near-stationary plasma (inter-ELM) are interrupted by short quasi-periodic bursts of intense plasma fluxes (intra-ELM). Characteristic inter-ELM heat loads had a peak at $\sim80$\,MW/m$^2$, while typical intra-ELM heat fluxes reached $\sim700$\,MW/m$^2$. The ELM frequency was $\sim 70$\,Hz and the ELM duration was $\sim3$\,ms. Note that ELM heat loads, stemming from edge plasma instabilities, exhibit strong spatiotemporal variations~\cite{Zohm1996,Leonard2014}. Thus, these values are merely indicative.

\subsection{Simulation setup}

The simulation domain is presented in Fig.\ref{fig:ir_domain}. The grids consist of 3 levels. The finest grid (level 2) extends for a minimum distance $400\,\mu$m from the free surface and the level 1 grid extends for $800\,\mu$m. These are chosen by considering the Ir thermal diffusivity at the melting point with the typical ELM duration as a time scale, which results in a characteristic heat diffusion length of $200\,\mu$m. Empirically, the finest AMReX level should be few times larger than the characteristic heat diffusion length, a rule of thumb that can easily be checked with convergence tests. Inaccuracies due to overly aggressive re-gridding are introduced on the solution of the highest level only through the boundary condition applied at the fine-coarse interface. The refinement ratio is two between levels 0 and 1 and four between levels 1 and 2. The discretization $\Delta{y}$ is chosen to ensure that the temperature gradients within the melt layer are well-resolved. The tangential discretizations $\Delta x$ and $\Delta z$ can be typically much larger than $\Delta y$ owing to the peculiarity of molten pools in fusion scenarios which render the shallow-water approximation highly accurate.

\begin{figure}
    \centering
    \includegraphics[width=\columnwidth]{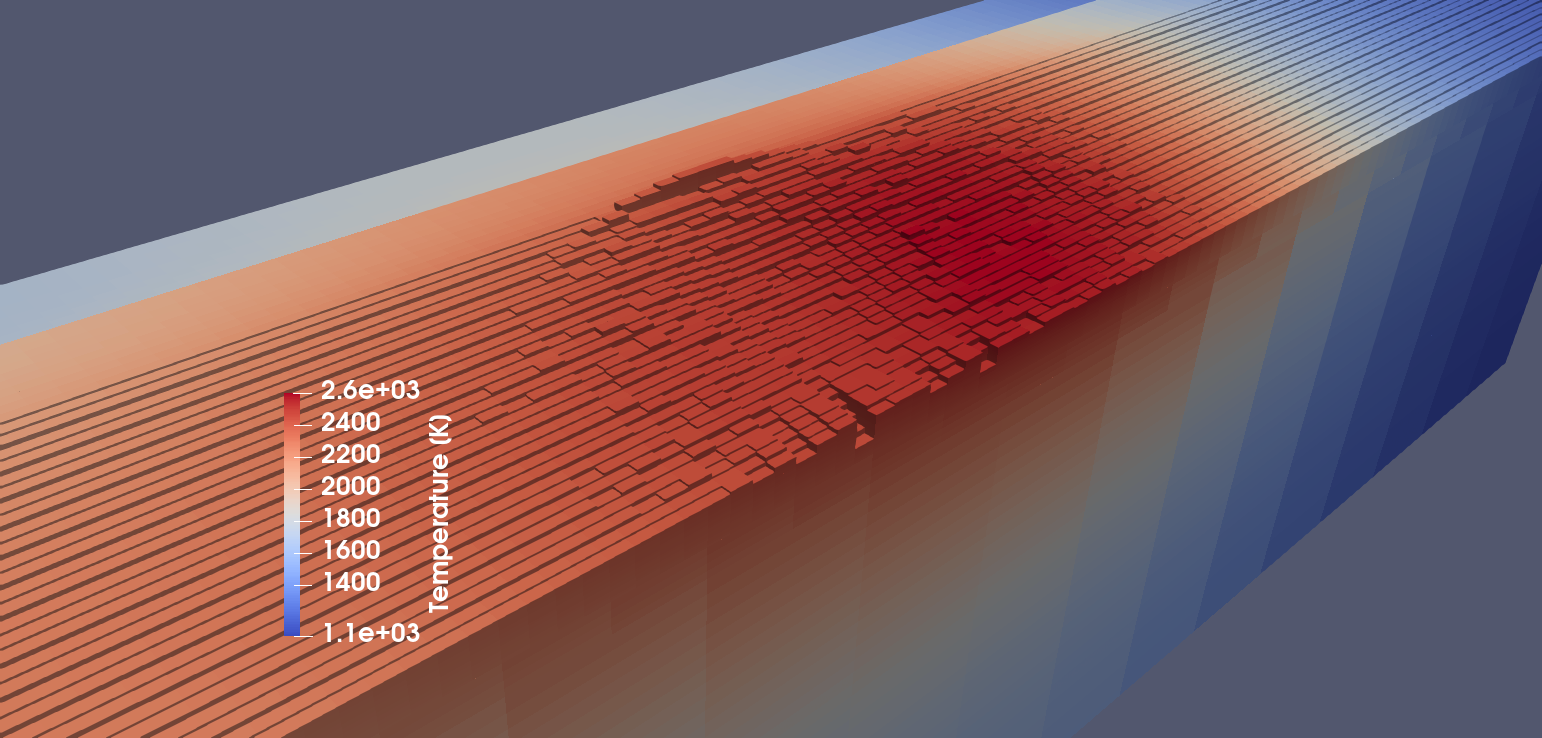}
    \caption{The 3D temperature profile on the deformed domain $4\,$s after the start of the simulation.}
    \label{fig:heat_output_3d}
\end{figure}

The sloped surface forms a 15$^\circ$ angle with the {\it zx } plane. Referring to the surfaces introduced in Eqs.\eqref{eqn:heat_BC}-\eqref{eqn:ground}, the $y=0$ plane corresponds to $S_{\mathrm{g}}$, the orange outlined surface to $b_2$, and the remaining surfaces to $S_{\mathrm{i}}$. For the heat solver, $q_{\mathrm{ext}}$ is zero on all surfaces except on $b_2$.

As far as the choice of the time step, when the explicit heat solver is used, the diffusion term often sets the most stringent stability criterion. The advection terms and the surface tension term in the momentum equation might also become unstable and could require extra considerations. Here, since the explicit heat solver is employed, the time step in level 0 was set to $40\,\mu$s for the stability of the diffusion term. Sub-cycling has also been enabled.

In this particular simulated case, the surface tension term in Eq.\eqref{eqn:momentum_simple} has been disabled in order to avoid the loss of melt from the edges. The reason is discussed in section \ref{section:Discussion}. The capping heights in the viscous damping and the Marangoni terms are set to $10\,\mu$m and to $5\,\mu$m respectively, well below the melt pool depths typically encountered in these simulations.

\subsection{Simulation outcome}

The intermittent nature of the incident heat loads of ELMing H-mode plasmas implies that the initial melt pools are transient, \emph{i.e.}, shallow melting is achieved within the intra-ELM period and complete resolidification takes place within the inter-ELM period. Even for the energetic AUG ELMs, the first instance of transient melting  requires that the inter-ELM sample temperature is sufficiently close to the Ir melting point. Close to the end of the exposure, after $\sim3.5$\,s, melt pools become sustained as they no longer fully resolidify between ELMs. Pools are displaced by the momentum sources contained in Eq.\eqref{eqn:momentum} with the Lorentz force typically being by far the dominant. However, in this particular experiment, due to the iridium material properties, the Marangoni flow and gravity were also important sources of momentum.

\begin{figure}
    \centering
    \includegraphics[width=\columnwidth]{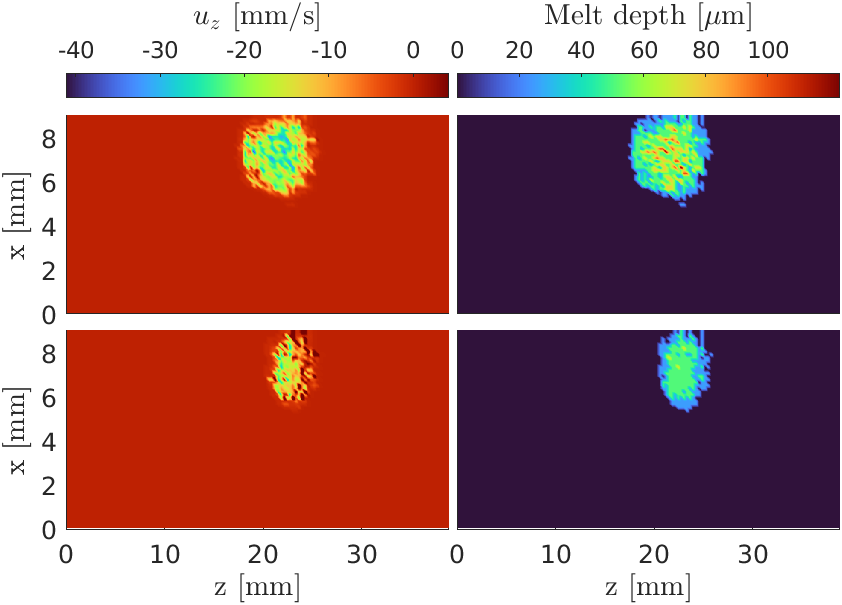}
    \caption{The melt depth and the velocity component (along gravity and the Lorentz force) at two different time instants, 3.79\,s (top) and 3.89\,s (bottom).}
    \label{fig:output_2d}
\end{figure}

The melt pools are continuously displaced. Thus, the surface deformation is building up during every transient melting event. The temperature field on the evolving 3D domain is shown in Fig.\ref{fig:heat_output_3d}. This output refers to the temperature at the center of all the cells, generated 4\,s into the simulation.

\begin{figure}[t]
    \centering
    \includegraphics[width=0.89\columnwidth]{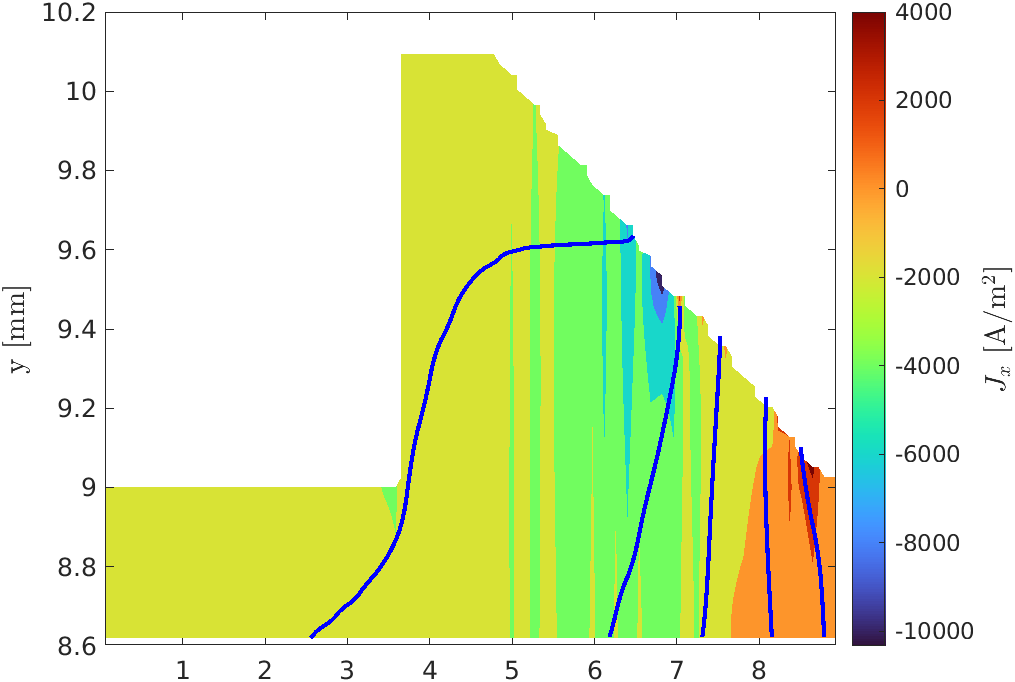}
    \includegraphics[width=0.89\columnwidth]{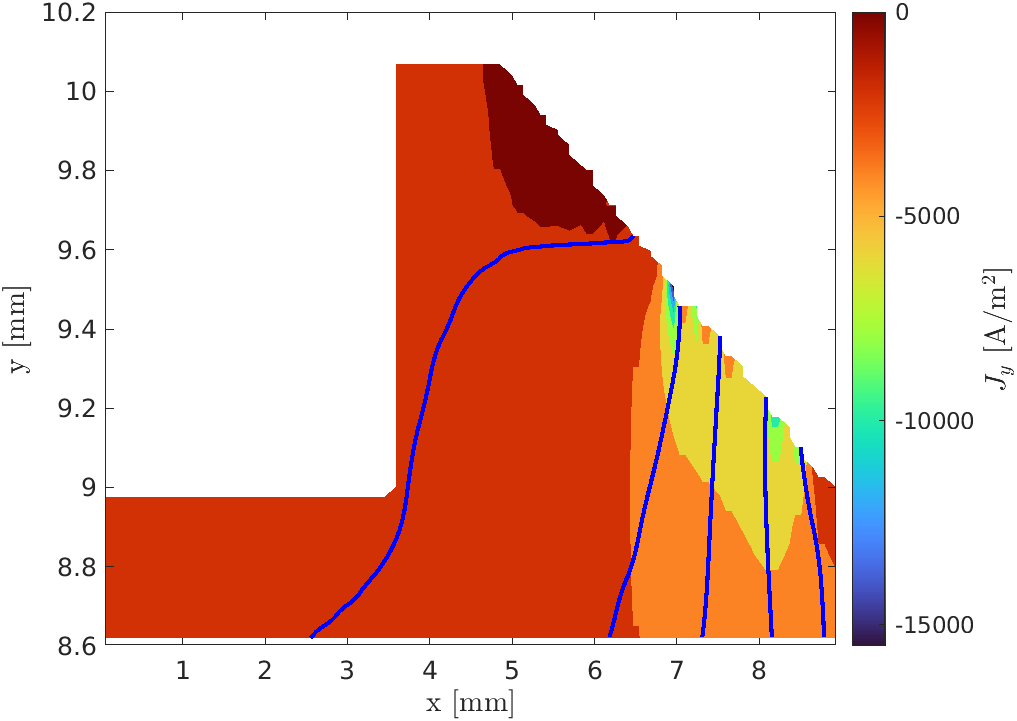}
    \caption{The replacement current density and its streamlines flowing through the sample at the 3.12\,s time instant.}
    \label{fig:potential_3d_slice}
\end{figure}

The 2D data can be post-treated to visualize the domain deformation, instantaneous melt depth and depth-averaged velocity field. An example is presented in Fig.\ref{fig:output_2d} for two different time instants. The illustrated melt pools are taken at time instants that belong to the sustained melting part of the simulation. The first instant is shortly after an ELM strikes, while the other instant is within the inter-ELM period (in the course of the incomplete re-solidification).

The current density is shown in Fig.\ref{fig:potential_3d_slice} for a {\it xy}-plane cut through the $z=0.021$\,m position at 3.12\,s after the start of the simulation. Here, the magnetic flux density is aligned with $\hat{\boldsymbol{x}}$ and hence only $J_y$ contributes to the Lorentz force. As expected from the grounded surface corresponding to $y=0$ and the emitting pristine surface forming $15^\circ$ angle with the {\it xz} plane, a rather weak bending of the current streamlines is observed.

\begin{figure}
    \centering
    \includegraphics[width=0.89\columnwidth]{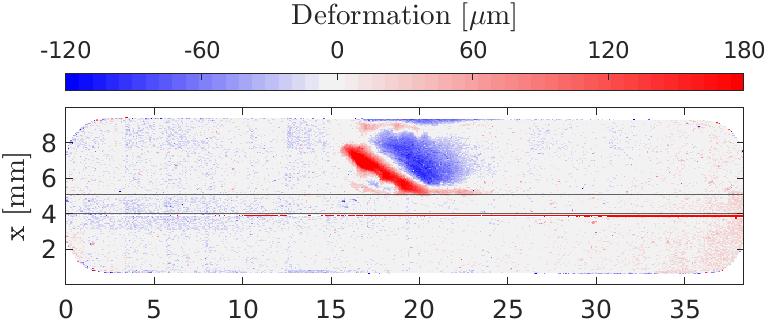}
    \includegraphics[width=0.89\columnwidth]{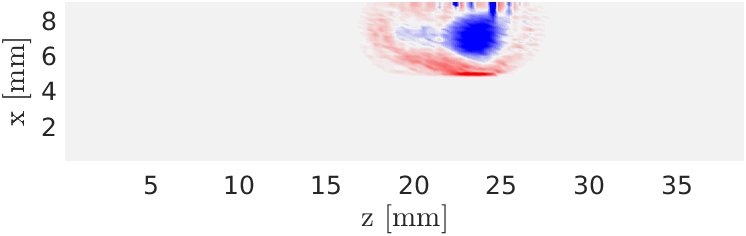}
    \caption{The final deformation profile in the experiment (top) and the simulation (bottom).}
    \label{fig:deformation_pofile}
\end{figure}

Finally, the deformation profile is a key output of the simulation, since it can be directly compared to the experiment. Such a comparison is presented in Fig.\ref{fig:deformation_pofile} revealing that MEMENTO is capable to reproduce the post-mortem evidence both qualitatively and quantitatively. Note that the shift between the MEMENTO deformation profile and the observations is well within the experimental uncertainties on the position of the heat-flux on the tile ($<1$ cm). For more detailed discussions on the melt dynamics and the final deformation, the reader is addressed to Ref.\cite{Ratynskaia2022_1}. Once again, it is pointed out that deformation is accumulated from all discrete melting events and that resolidification takes place in between melting.

\section{Summary and discussion}\label{section:Discussion}

The status of MEMENTO's solvers will be briefly recapped to point out possible future updates. Furthermore, inherent limitations of the code in specific fusion-relevant applications due to the underlying shallow water assumption, will be discussed.

The \emph{heat solvers} provide solutions on the deformed domain and, apart from diffusion, also handle convection as well as phase transitions. The solvers also enable heat transfer simulations under volumetric loading relevant for plasma-facing component interaction with relativistic runaway electrons.
Simulations of gaps are also feasible by introducing regions with zero thermal conductivity.

The \emph{electrostatics solver} enables solutions of the replacement current propagation every time step, which leads to more accurate computation of (the often dominating) Lorentz force. The solution is found in the deformed domain and includes the dynamo term, which is relevant to future reactor scenarios. This is the most computationally heavy module of the code. Depending on the desired accuracy and speed, the solver can be utilized to provide an approximate solution for the melt depth-averaged current through a reduction coefficient to the escaping emitted current density value. Such a solution can be then utilized for the specific choice of geometry and enable sufficiently accurate runs employing only the heat and fluid solvers.

The \emph{fluid solver} provides the solution to the shallow waters equations. Depending on the scenario, to address stability issues inherent to the shallow water model, it might require truncation of the melt depth column to some minimal values, set by the input. The introduction of such melt depth thresholds can be entirely avoided or set to negligible values by including the surface tension in the simulations; an optimal choice both from a physics and a numerical point of view. However, in scenarios where the melt pool is created at or moved to an edge, the shallow waters approximation cannot capture the 3D nature of the resulting pool curvature and the details of the corner wetting, which is a limitation of the physics model rather than of the implementation. In case the scenario features an insignificant forcing of the pool off the edge, then an impermeable wall can provide an adequate solution. Such a solution was employed in the modelling of a recent AUG melting experiment, where all features of the experimental surface deformation were successfully recovered in the MEMENTO runs~\cite{Ratynskaia2023}. Future optimization of this module primarily concerns the surface tension implementation. In particular, other numerical schemes will be investigated in an attempt to improve the flexibility with respect to the choice of grid resolution.

The current version of MEMENTO is capable of modelling most fusion-relevant plasma-facing component layouts and plasma wetting geometries, including reactor designs with cooling pipes while resolving the ultra-thin melt layer dynamics. MEMENTO also efficiently addresses temporal scale separations and allows simulations of diverse plasma loading scenarios, from near-stationary L-mode plasmas, ELM-free H-mode regimes and quasi-periodic ELMing H-mode discharges to plasma disruption thermal or current quenches and runaway electron termination. On the computational front, the remaining optimization issue to address concerns parallelization with MPI to facilitate MEMENTO simulations in distributed memory machines.

\begin{figure}[t]
    \centering
    \begin{overpic}[width=0.27\linewidth]{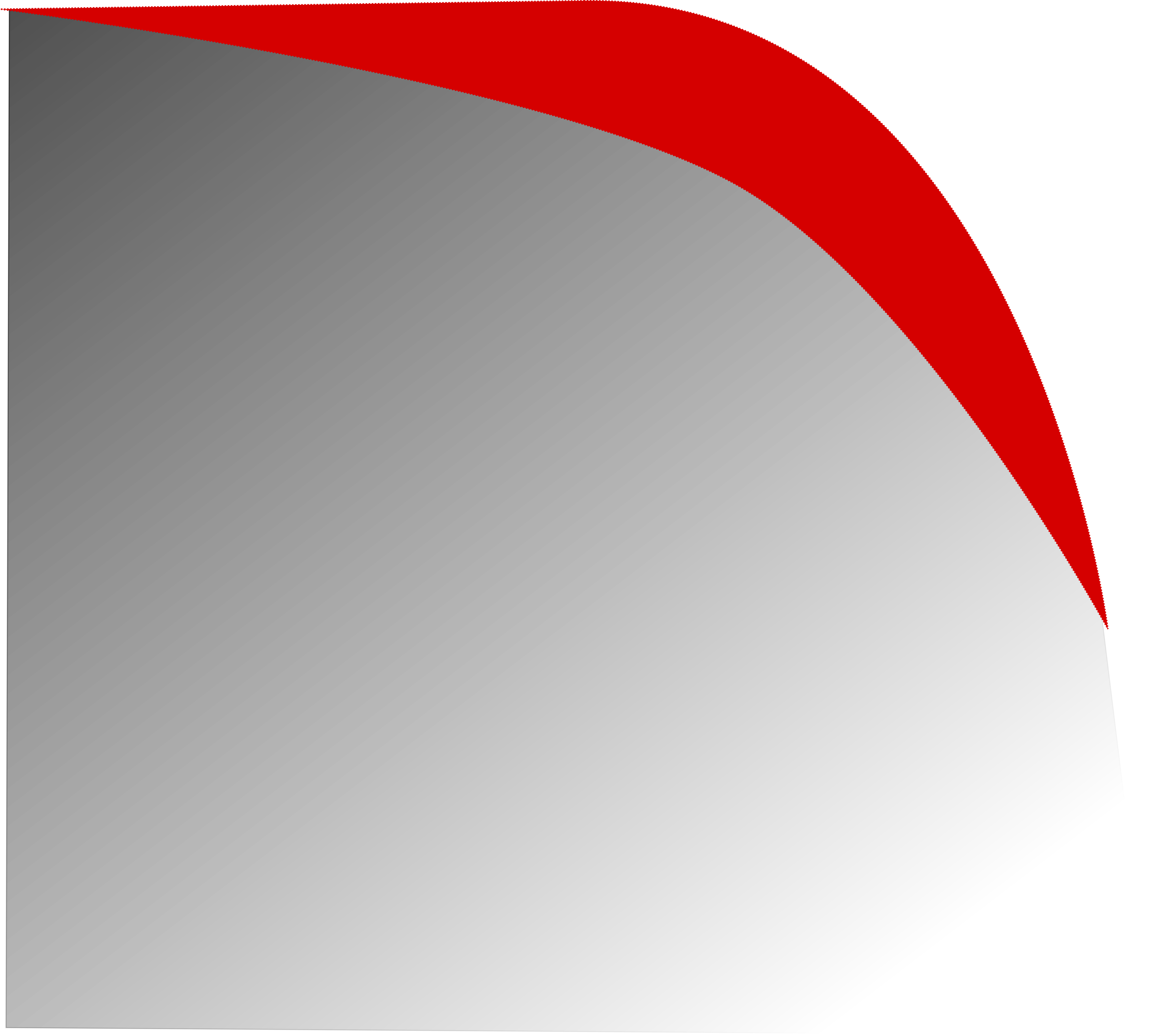}
      \put(10,62){\textcolor{SkyBlue}{\textbf{(a)}}}
    \end{overpic}
    \begin{overpic}[width=0.24\linewidth]{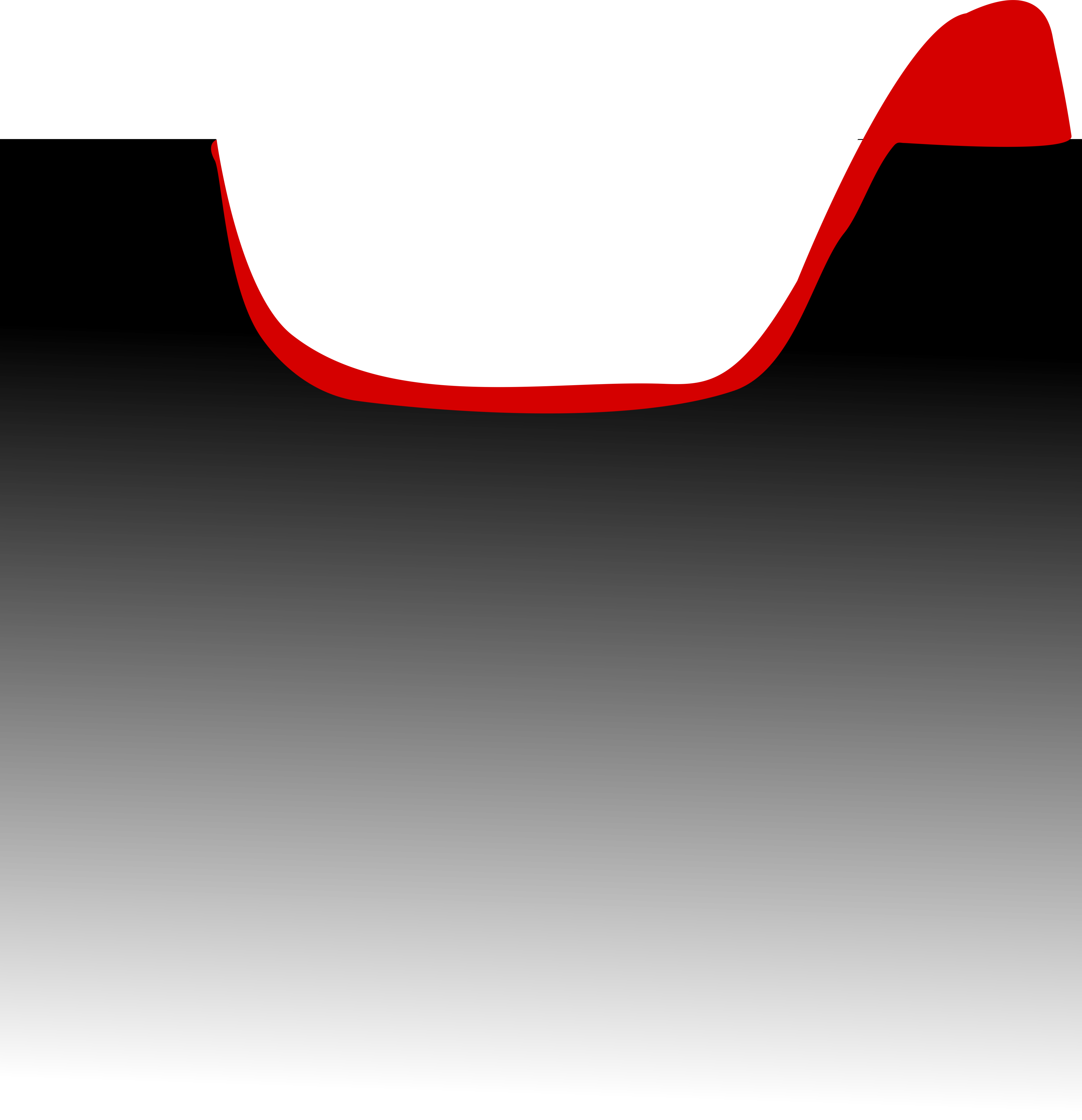}
      \put(10,53){\textcolor{SkyBlue}{\textbf{(b)}}}
    \end{overpic}
    \begin{overpic}[width=0.24\linewidth]{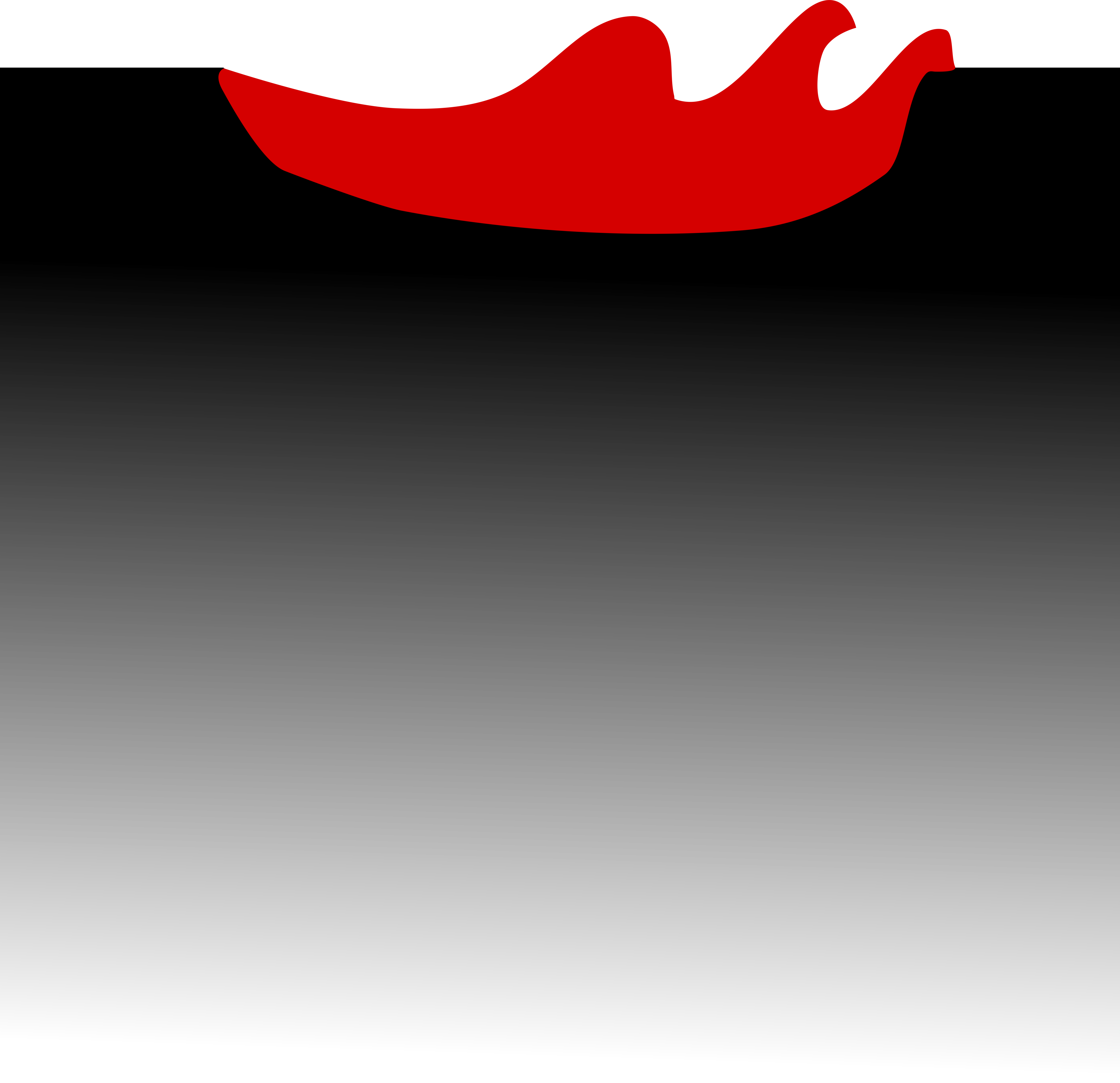}
      \put(10,62){\textcolor{SkyBlue}{\textbf{(c)}}}
    \end{overpic}
    \caption{Sketches of scenarios in which MEMENTO's implementation of the shallow water approximation breaks down. The red color corresponds to the molten material while the gray color illustrates the solid. The local surface normal vectors form an angle near $45^\circ$ due to (a) the geometry of the initial free surface, (b) excessive melting and deformation. (c) The free surface position is not assigned a unique value for any tangential coordinate pair.}
    \label{fig:limitations}
\end{figure}

Finally, it is stressed again that the shallow water approximation constitutes the major reason for MEMENTO's ability to reliably model fusion-relevant melting scenarios where meter-long plasma-wetted areas must be simulated while resolving micrometer-deep melt pools. Simultaneously, MEMENTO's implementation of shallow water averaging imposes constraints on certain types of scenarios. There are two associated limitations. The first limitation concerns cases where the free surface includes a curvature or an angle such that the local surface normal vectors (at positions where melting is realized) form an angle which is near or above $45^{\circ}$ degrees. This might occur because such features are either inherent to the initial free surface, as in Fig.\ref{fig:limitations}a, or generated when excessive melting and forcing result in drastic material excavations, as in Fig.\ref{fig:limitations}b. The second limitation concerns the fact that each pair of the tangential coordinates, i.e. the $x$ and $z$ directions for MEMENTO as illustrated in Fig.\ref{fig:MEMENTO_coordinate_sketch}, should uniquely specify one value of the free surface position $y$. This means that wave breaking phenomena, as sketched in Fig.\ref{fig:limitations}c, cannot be modelled.

Since the shallow water originating limitations mainly concern the aforementioned two cases, it can be stated that the majority of complex fusion-relevant melting scenarios can be tackled with the MEMENTO simulations. Moreover, in order to facilitate the self-consistent modelling of small-scale melt dynamics and melt flow stability in fusion reactors, the MEMENTO simulation outcome can be utilized in specialized ANSYS CFD set-ups. Such a work-flow enables the faithful reproduction of the macroscopic picture (large wetted areas, three-dimensional heat transfer, long time thermal response, large scale melt dynamics) responsible for the eventual melt pool depths $h$ and velocities $u$, while allowing for full detail simulations of a small domain of interest where $h$ and $u$ are imposed through boundary and initial conditions~\cite{Vignitchouk2022}. After the necessary tests against experimental evidence from melting events in the JET tokamak~\cite{Vignitchouk2022}, this work-flow has been employed for predictive modelling of melt splashing in ITER~\cite{Vignitchouk2023} and will be utilized in predictive studies for DEMO.

\appendix
\section{Simulation flowchart}\label{sec:appendix}
\setcounter{figure}{0}

The flowchart of a MEMENTO simulation is shown in Fig.\ref{fig:flowchart}. In short, a simulation starts by reading the input and performing the necessary initializations. The free surface is generated at the specified position, the initial temperature is set, the melt thickness is initialized to zero and the material properties are calculated. If sub-cycling is not enabled, the fluid solver is called first, followed by the heat solver and the electrostatics solver. If sub-cycling is enabled, the order of operations changes. The heat equation is first solved on all levels except the highest. Then, the shallow waters, heat and electrostatics solvers are called in sequence and repeatedly according to the time step division by the refinement ratio. Once all solvers are updated, the code checks an input-defined frequency to determine whether output is to be generated in the current time step. Finally, the termination conditions are checked to see whether the simulation has reached its end. If not, time is incremented and the calculations are repeated.
\begin{figure*}
\centering
\resizebox{13cm}{!}{\begin{tikzpicture}[node distance=2cm]

\node (start) [startstop] {Start};

\node (in1) [process, below of=start, yshift=0.5cm] {Read input, initiliaze and calculate $\Delta$t for all levels};

\node (new_time) [process, below of=in1, yshift=0.5cm] {Start calculations for time t};

\node (sub_check) [decision, below of=new_time, yshift=-0.8cm] {Subcycling in time};

\node (regrid_all) [process, below of=sub_check, yshift=-1.0cm] {Regrid all levels, set $l=0$};

% \node (level0) [process, below of=regrid_all, yshift=0.5cm] {Start from $l=0$};

\node (AdvanceSW_ns) [process, below of=regrid_all, yshift=0.1cm] {If Eq.1 or Eq.2 are solved, advance SW of $\Delta t$ (update $b_2$ and $\boldsymbol{U}$)};

\node (Reval_ns) [process, below of=AdvanceSW_ns, yshift=-0.0cm] {Revaluate the heat domain at level $l$ (update $b_2$ in the heat domain)};

\node (Advance_HT_ns) [process, below of=Reval_ns, yshift=0.2cm] {If Eq.3 is solved advance heat transfer by $\Delta$t at level $l$ (update $T$ and $b_1$)};

\node (level_check) [decision, below of=Advance_HT_ns, yshift=+0.0cm] {$l=l_{max}$};

\node (Advance_electro_ns) [process, below of=level_check, yshift=-0.5cm] {If melt exists, Eq.4 and Eq.5 are solved and $l=l_{\mathrm{max}}$ compute $\psi$  on all levels and $\langle \boldsymbol{J} \rangle$ on $l_{\mathrm{max}}$ (update $\psi$ and $\langle \boldsymbol{J} \rangle$)};

\node (incr_level) [process_mini, left of=level_check, xshift=-0.4cm] {$l=l+1$};

\node (synch_lvls) [process, below of=Advance_electro_ns, yshift=-0.0cm] {Synchronize levels};

% \node (electro) [process, below of=synch_lvls, yshift=-0.5cm] {Solve electrostatics (update $\boldsymbol{J}$)};

\node (set_l) [process, right of=sub_check, xshift=4.0cm] {Set $l=0$, $c_{0,1,..,l_{\mathrm{max}}}=0$};

\node (regrid_one) [process, below of=set_l, yshift=0.5cm] {Regrid level $l$};
% \node (regrid_one) [process, below of=sub_check, yshift=0cm] {Regrid level $l=0$};
\node (AdvanceSW_s) [process, below of=regrid_one, yshift=0.2cm] {If $l = l_{max}$ and Eq.1 or Eq.2 are solved, advance SW by $\Delta \mathrm{t}/(r_1 \cdot r_2 \cdot ... \cdot r_l)$ (update $b_2$ and $\boldsymbol{U}$)};

\node (Reval_s) [process, below of=AdvanceSW_s, yshift=0cm] {Revaluate the heat domain at level $l$ (update $b_2$ in the heat domain)};

\node (Advance_HT_s) [process, below of=Reval_s, yshift=0.0cm] {If Eq.3 is solved advance heat transfer by $\Delta \mathrm{t}/(r_1 \cdot r_2 \cdot ... \cdot r_l)$ at level $l$ (update $T$ and $b_1$)};

\node (Advance_electro_s) [process, below of=Advance_HT_s, yshift=-0.4cm] {If melt exists, Eq.4 and Eq.5 are solved and $l=l_{\mathrm{max}}$ compute $\psi$  on all levels and $\langle \boldsymbol{J} \rangle$ on $l_{\mathrm{max}}$ (update $\psi$ and $\langle \boldsymbol{J} \rangle$)};

\node (level_check_s) [decision, below of=Advance_electro_s, yshift=-0.5cm] {$l=l_{\mathrm{max}}$};

\node (incr_level_s) [process_mini, right of=level_check_s, xshift=0.7cm] {$c_l=c_l+1$ $l=l+1$};

\node (recur_check_s) [decision, below of=level_check_s, yshift=-0.5cm] {$c_l=r_l$};

\node (incr_cnt) [process_mini, left of=recur_check_s, xshift=-0.5cm] {$c_l=c_l+1$};

\node (synch_s) [process, below of=recur_check_s, xshift=0.0cm] {Synchronize all levels $l^\prime$ for which $c_{l^\prime+1}=r_{l^\prime+1}$, set $c_{l^\prime+1}=0$, $l=\mathrm{min}({l^\prime})$};

\node (break_check_s) [decision, right of=synch_s, xshift=2.5cm] {$l=0$};

% \node (recurs) [recursive, right of=level_check_s, xshift=3.5cm] {Advance level 1 $r_1$ times, where $r_1$ is the refinement ratio between levels 0 and 1};

% \node (sync01) [process, below of=recurs, yshift=-0.0cm] {$l_{max} = 0$};

\node (output) [process, below of=synch_s, xshift=-2.9cm] {Write output files};

\node (end_check) [decision, below of=output] {$t>t_{max}$};

\node (incr_t) [process_mini, left of=end_check, xshift=-4.5cm] {$t = t + \Delta t$};

\node (end) [startstop, right of=end_check, xshift=1.0cm] {End simulation};

\draw [arrow] (start) -- (in1);
\draw [arrow] (in1) -- (new_time);
\draw [arrow] (new_time) -- (sub_check);
\draw [arrow] (sub_check) -- node[anchor=south] {yes} (set_l);

\draw [arrow] (sub_check) -- node[anchor=west] {no} (regrid_all);
\draw [arrow] (regrid_all) -- (AdvanceSW_ns);
\draw [arrow] (AdvanceSW_ns) -- (Reval_ns);
\draw [arrow] (Reval_ns) -- (Advance_HT_ns);
\draw [arrow] (Advance_HT_ns) -- (level_check);
\draw [arrow] (level_check) -- node[anchor=south] {no} (incr_level);
\draw [arrow] (level_check) -- node[anchor=west] {yes} (Advance_electro_ns);
\draw [arrow] (Advance_electro_ns) -- (synch_lvls);
\draw [arrow] (incr_level) |- (AdvanceSW_ns);
\draw [arrow] (incr_level) |- (AdvanceSW_ns);
\draw [arrow] (synch_lvls) |- (output);

\draw [arrow] (set_l) -- (regrid_one);
\draw [arrow] (regrid_one) -- (AdvanceSW_s);
\draw [arrow] (AdvanceSW_s) -- (Reval_s);
\draw [arrow] (Reval_s) -- (Advance_HT_s);
\draw [arrow] (Advance_HT_s) -- (Advance_electro_s);
\draw [arrow] (Advance_electro_s) -- (level_check_s);
\draw [arrow] (level_check_s) -- node[anchor=south] {no} (incr_level_s);
\draw [arrow] (incr_level_s) |- (regrid_one);
\draw [arrow] (level_check_s) -- node[anchor=west] {yes} (recur_check_s);
\draw [arrow] (recur_check_s) -- node[anchor=south] {no} (incr_cnt);
\draw [arrow] (incr_cnt) |- (regrid_one);
\draw [arrow] (recur_check_s) -- node[anchor=west] {yes} (synch_s);
\draw [arrow] (synch_s) -- (break_check_s);
\draw [arrow] (break_check_s) |- node[anchor=north] {yes} (output);
\draw [arrow] (break_check_s) |- node[anchor=south] {no} (regrid_one);

\draw [arrow] (output) -- (end_check);
\draw [arrow] (end_check) -- node[anchor=south] {no} (incr_t);
\draw [arrow] (incr_t) |- (sub_check);
\draw [arrow] (end_check) -- node[anchor=south] {yes} (end);

\end{tikzpicture}

% \end{document}
}
\caption{Flowchart of a MEMENTO simulation. Here $r_l$ is the refinement ratio of level $l$. It is pointed out that, in MEMENTO, subcycling is implemented with a recursive algorithm.}\label{fig:flowchart}
\end{figure*}

\section*{Acknowledgments}

\noindent SR \& PT acknowledge the financial support of the Swedish Research Council under Grant No\,2021-05649. The work has also been performed within the framework of the EUROfusion Consortium,\,funded by the European Union via the Euratom Research and Training Programme (Grant Agreement No\,101052200 - EUROfusion). The views and opinions expressed are however those of the authors only and do not necessarily reflect those of the European Union or the European Commission. The European Union or European Commission cannot be held responsible for them. The MEMENTO simulations were enabled by resources provided by the National Academic Infrastructure for Supercomputing in Sweden (NAISS) and the Swedish National Infrastructure for Computing (SNIC) at the NSC (Link\"oping University) partially funded by the Swedish Research Council through grant agreements No\,2022-06725 and No\,2018-05973.

%\section*{Conflict of interest statement }
% On behalf of all authors, the corresponding author states that there is no conflict of interest.

\clearpage

\bibliography{biblio_new}

\end{document}